\pdfoutput=1
\documentclass[a4paper,fleqn]{cas-sc}

\usepackage[numbers]{natbib}
\usepackage{amsmath, amssymb, amsfonts}
\usepackage{graphicx}
\usepackage{color}

\def\calO{\mathcal{O}}

\usepackage{xspace}
\newcommand{\dfem}{$\partial$FEM\xspace}
\definecolor{code}{rgb}{0.7, 0, 0.4}
\newcommand{\code}[1]{\texttt{\small\color{code} #1}}

\begin{document}

\shorttitle{Implicit Lagrangian Hydrodynamics}
\shortauthors{Andrej et~al.}

\title [mode = title]{Implicit Lagrangian Hydrodynamics \texorpdfstring{\\}{ }
                      with High-Order Finite Elements}

\author[aff1]{Julian Andrej}
\author[aff1]{John Camier}
\author[aff1]{Veselin Dobrev}
\author[aff1]{Tzanio Kolev}
\author[aff1]{Boyan Lazarov}
\author[aff1]{Ketan Mittal}
\author[aff1]{Robert Rieben}
\author[aff1]{Brandon Talamini}
\author[aff1]{Vladimir Z. Tomov}\cormark[1]

\affiliation[aff1]
{organization={Lawrence Livermore National Laboratory},
 city={Livermore}, state={CA}, country={USA}}

\cortext[1]{Corresponding author, tomov2@llnl.gov}

\nonumnote{Performed under the auspices of the U.S. Department of Energy under
Contract DE-AC52-07NA27344 (LLNL-JRNL-2023211).}

\begin{abstract}
We present an implicit time integration capability for
high-order curvilinear finite element Lagrangian hydrodynamics.
Starting from an existing explicit formulation, the implicit treatment builds
directly on the original discretization and operator structure and does not
alter the underlying spatial formulation or physics model.
We demonstrate our implementation using the Laghos miniapp \cite{Laghos},
which is built on the MFEM finite element library \cite{MFEM,MFEM2020,MFEM2024}.
To support gradient-based nonlinear solution methods, we compute Jacobian
actions automatically using MFEM's \dfem interface together with
Enzyme-based automatic differentiation, and apply the resulting Jacobian in a
matrix-free or fully-assembled manner within a Newton-Krylov solver.
To ensure robust and differentiable nonlinear solves in the presence of shocks,
we introduce a smooth artificial viscosity treatment based on smooth
approximations of non-differentiable pointwise operations. The
differentiable artificial viscosity  presently lacks a limiter to
ensure high-order scaling away from shocks, but is sufficient for illustrating
the benefits of implicit Lagrangian hydrodynamics.
The behavior and performance of the implicit method are demonstrated
on several standard benchmark problems.
We verify high-order convergence on the smooth Taylor-Green vortex in the absence of
artificial viscosity,
show correct strong-shock behavior on the Sedov blast problem, and obtain
significant improvements in accuracy-per-time-to-solution on the Triple Point
problem where explicit stability constraints become increasingly severe for
high-order discretizations.
\end{abstract}

\maketitle

%-------------------------------------------------

\section{Introduction}
\label{sec_intro}

Implicit methods in Lagrangian hydrodynamics simulations have gained substantial
attention due to their inherent stability advantages, particularly their
ability to circumvent stringent Courant-Friedrichs-Lewy (CFL) conditions
characteristic of explicit schemes.
By enabling significantly larger time steps, implicit schemes offer efficiency
benefits in stiff, multi-scale, and long-duration hydrodynamics simulations.
Recent advances exemplified by Chiocchetti et al.~\cite{Russo2024}, for the
context of 1D acoustic metamaterial simulations,
and by Del Pino et al.~\cite{Despres2023, Despres2024}, targeting well-posed
first-order general Lagrangian hydrodynamics schemes in any dimension,
underscore these advantages through robust finite-volume implicit formulations
that achieve stable, conservative, and thermodynamically consistent solutions.
However, these existing methods predominantly rely on low-order
discretizations, typically first- or at most second-order,
thus limiting their resolution and accuracy.

High-order numerical methods, and high-order finite elements in particular,
have become increasingly attractive due to their ability to simultaneously
improve solution accuracy and computational efficiency.
By representing curved geometries and solution fields more accurately,
high-order discretizations can achieve a desired accuracy with substantially
fewer degrees of freedom than low-order methods.
They are also essential for attaining optimal convergence rates on domains
with curved boundaries and interfaces, as well as for preserving symmetry in
radially symmetric flows \cite{Fisher2002,Orszag1979}.
These properties are particularly important in Lagrangian hydrodynamics,
where the mesh moves and deforms with the flow, requiring
geometric fidelity to be maintained throughout the simulation \cite{Dobrev2012}.
Although high-order methods increase the amount of computation
performed per degree of freedom, this additional arithmetic intensity
maps well to modern GPU computing architectures and often leads to
improved overall performance \cite{MAPP2020,CEED2021}.

In this work, we introduce a novel implicit Lagrangian hydrodynamics method
leveraging automatic differentiation (AD) \cite{GriewankWalther2008}
within our existing high-order
finite element framework \cite{Dobrev2012}.
This work continues our broader effort to incorporate AD into high-order finite
element methods and related nonlinear optimization-based simulations
\cite{MFEM2024, DIFF2025, TMOP2026, Andrej2026}.
By utilizing AD, we keep the original discretization method structurally
unaltered, seamlessly extending its capabilities into the implicit domain.
The proposed method represents the first implicit formulation based on a
finite element discretization, enabling significant improvement in
computational efficiency and resolution.
The approach achieves high-order spatial and temporal convergence in smooth
regions (in the absence of artificial viscosity), providing improved resolution 
compared to existing implicit methods.
The Jacobian of the system is computed in a matrix-free manner using
partial assembly (PA), requiring only action-based evaluations,
which notably optimizes computational resources. Fully-assembled Jacobians are
also supported when needed, e.g. when utilizing direct linear solvers.
We demonstrate that a single unified code, via the Laghos miniapp,
can run efficiently on both CPUs and GPUs,
preserving all the capabilities and flexibility of the
original formulation \cite{Dobrev2012}, thus substantially broadening
the spectrum of the implicitly tractable hydrodynamics problems.

%-------------------------------------------------

\section{Preliminaries}
\label{sec_prelims}

In this section we describe the system we consider, its discretization in space
and time, and the notation we utilize.
We also discuss its efficient implementation based on finite element partial
assembly and applicability to GPU computing.

%---------------------

\subsection{Compressible Euler equations and FE discretization}
The nonconservative form of the compressible Euler equations of gas dynamics in
a Lagrangian reference frame is given by the following system:
\begin{align}
	\label{eq_rho_cont}
	\text{Mass Conservation:} \quad
	\frac{1}{\rho} \frac{d \rho}{d t} & = - \nabla \cdot v,    \\
	\label{eq_v_cont}
	\text{Momentum Conservation:} \quad
	\rho \frac{d v}{d t}              & = \nabla \cdot \sigma, \\
	\label{eq_e_cont}
	\text{Specific Internal Energy Evolution:} \quad
	\rho \frac{d e}{d t}              & = \sigma : \nabla v,   \\
	\label{eq_x_cont}
	\text{Equation of Motion:} ~ \quad
	\frac{d x}{d t}                   & = v,                   \\
	\text{Equation of State:} ~~~ \quad
	\sigma                            & = - EOS(\rho, e) I,
\end{align}
where the independent variables $\rho, v, e, x$ represent material density,
material velocity, specific internal energy, and particle position, respectively.
The general stress tensor $\sigma$ is computed from the equation of state
for the pressure $p = p(\rho, e)$ and
includes artificial viscosity stresses needed to regularize the system.
All time derivatives on the left-hand sides are material derivatives, namely,
$d \rho / dt = \partial \rho / \partial t + v \cdot \nabla \rho$.

Our spatial discretization is based on a mixed finite element method,
originally introduced in~\cite{Dobrev2012}.
Referring to $\rho, v, e, x$ as the discrete fields of interest, we have
$v \in \mathcal{V} \subset [H^1(\Omega_0)]^d$
with basis $\{w_i\}_1^{N_\mathcal{V}}$,
and $e \in \mathcal{E} \subset L_2(\Omega_0)$
with basis $\{\phi_i\}_1^{N_\mathcal{E}}$,
where $d$ is the spatial dimension and $\Omega_0$ is the initial domain:
\[
	v(x_0, t) = \sum_{i = 1}^{N_\mathcal{V}} v_i(t) w_i(x_0)
	= \mathbf{v}^\top \mathbf{w}, \quad
	e(x_0, t) = \sum_{j = 1}^{N_\mathcal{E}} e_j(t) \phi_j(x_0)
	= \mathbf{e}^\top \boldsymbol{\phi},
\]
Here $\mathbf{v}$ and $\mathbf{e}$ are
the unknown time-dependent solution vectors.
Material positions are discretized using the expansion
\[
	x(x_0, t) = \sum_{i = 1}^{N_\mathcal{V}} x_i(t) w_i(x_0)
	= \mathbf{x}^\top \mathbf{w},
\]
where the vector $\mathbf{x}$ stores the positions
of the nodes of $\mathcal{V}$ at time $t$.
When specifying polynomial degrees for these spaces, we use the shorthand Q$k$Q$l$,
where the first entry, $k$, denotes the degree of the continuous kinematic space
$\mathcal{V}$ used for $x$ and $v$, and the second, $l$, denotes the
degree of the discontinuous thermodynamic space $\mathcal{E}$ used for $e$.
The weak forms of~\eqref{eq_v_cont} and~\eqref{eq_e_cont} become
\begin{subequations}
\label{eqs_semi}
	\begin{align}
		\label{eq_v_semi}
		\mathbf{M}_\mathcal{V} \frac{d \mathbf{v}}{dt} & = - \mathbf{F} \cdot \mathbf{1},      \\
		\label{eq_e_semi}
		\mathbf{M}_\mathcal{E} \frac{d \mathbf{e}}{dt} & = \mathbf{F}^\top \cdot \mathbf{v},
	\end{align}
\end{subequations}
where
\begin{equation}
\label{eq_matrices}
	\mathbf{M}_\mathcal{V} =
	\int_{\Omega(t)} \rho \mathbf{w} \mathbf{w}^\top, \quad
	\mathbf{M}_\mathcal{E} =
	\int_{\Omega(t)} \rho \boldsymbol{\phi} \boldsymbol{\phi}^\top, \quad
	\mathbf{F}_{ij} = \int_{\Omega(t)} (\sigma : \nabla w_i) \phi_j,
\end{equation}
and $\mathbf{1}$ is a vector of ones of size $N_\mathcal{E}$.
Finally, the particle position values $x$ are evolved simply by~\eqref{eq_x_cont},
and the density values $\rho$ are evolved only at the quadrature points used to evaluate
the integrals in \eqref{eq_matrices}, thus retaining the notion of
pointwise mass conservation:
\[
	\frac{d \mathbf{x}}{d t} = \mathbf{v}, \quad
	\rho(x_0, t) = \frac{\rho_0(x_0)}{|J_{x \leftarrow x_0}(x_0, t)|},
\]
where $\rho_0$ is initial density, $J_{x \leftarrow x_0}=\nabla_{x_0}x$ is
the Jacobian of the transformation $x = x(x_0, t)$, and $|\cdot|$ denotes the
determinant.
% We also use $J_{x \leftarrow \hat{x}}$ and $J_{x_0 \leftarrow \hat{x}}$
% to represent Jacobians of transformations from the reference to
% physical and initial configurations, respectively.
The complete details of the above spatial discretization
can be found in~\cite{Dobrev2012}.
So far we have exclusively used explicit time integrators to evolve the system.
More recently, we have developed high-order accurate time integrators~
\cite{Sandu2021} that are customized to preserve total energy.

%---------------------

\subsection{Efficient implementation with PA}
\label{sec_pa}

\begin{figure}[pos=htbp]
  \centerline{\includegraphics[width=0.75\textwidth]{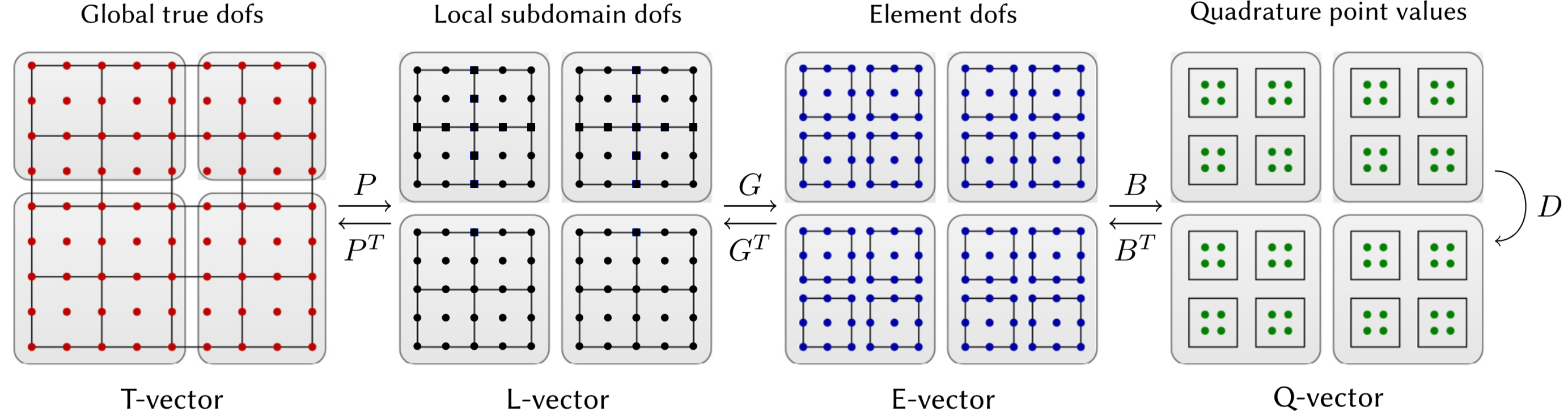}}
  \caption{Fundamental finite element operator decomposition.}
\label{fig_pgbd}
\end{figure}

Since this paper includes performance comparisons between explicit
and implicit approaches, it is useful to provide an overview
of their core implementation details.
Throughout this work we utilize the fundamental finite element (FE) operator
decomposition, namely, that any parallel FE operator
$A$ can be decomposed as:
\begin{equation}
	\label{eq_decompose}
	A = P^T G^T B^T D B G P,
\end{equation}
see Figure~\ref{fig_pgbd}, where
$P$ represents the subdomain restriction operator that maps the global MPI
vector (T-vector) to its local processor values (L-vector),
$G$ represents the element restriction operator that maps the L-vector to its
element values (E-vector),
$B$ represents the basis evaluator that interpolates an E-vector to the
quadrature points inside each element (Q-vector), and finally
$D$ is the operator that defines the quadrature-level computations.

In contrast to standard (full) finite element assembly, where the operator $A$
is computed and stored as a distributed sparse matrix, the
\textit{partial assembly} (PA) technique~\cite{Orszag1979,Fisher2002}
computes and stores only the result of $D$,
while evaluating the actions of $P, G$ and $B$ on-the-fly.
This approach significantly reduces data motion and eliminates the need for
global matrix storage, enabling highly efficient GPU execution.
Furthermore, PA takes advantage of the tensor product structure of the degrees
of freedom and quadrature points on quadrilateral and hexahedral elements to
perform the action of $B$ without storing it as a matrix; see specific details
in~\cite{MFEM2020,CEED2021,Maldonado2020}.
With these techniques, the storage, assembly, and application of the operator,
per mesh element, scale as
$\calO(p^{d})$, $\calO(p^{d})$, and $\calO(p^{d+1})$, respectively, compared to
$\calO(p^{2d})$, $\calO(p^{3d})$, and $\calO(p^{2d})$, for standard full assembly.

Our implementation of the techniques described above
for the case of equations \eqref{eq_v_semi} and \eqref{eq_e_semi}
is done in the Laghos miniapp \cite{Laghos}.
In the explicit time integration case, the primary computational kernels include
the application of the momentum and energy mass matrices
$\mathbf{M}_\mathcal{V}$ and $\mathbf{M}_\mathcal{E}$, whose quadrature data
is computed once for the whole simulation and reused throughout every conjugate
gradient (CG) solve with a diagonal Jacobi preconditioner,
and the evaluation of the quadrature data for the right-hand side operator
$\mathbf{F}$, which is assembled at each time integrator stage but applied
only twice per stage.
The major computational kernels for the implicit time integration case
are detailed in Section~\ref{sec_diff}.
The full implementation and open-source code are available in \cite{Laghos}.

%-------------------------------------------------

\section{Differentiation Approach}
\label{sec_diff}

This section describes the differentiation tools needed by the implicit
solver and the specifics of its efficient implementation in our finite element
setting.
By exploiting MFEM's partial assembly operator decomposition, differentiation is
localized to quadrature-point kernels, enabling both matrix-free and
fully-assembled Jacobian actions with minimal changes to the underlying
discretization.
We then summarize the Enzyme-based automatic differentiation workflow used in
our experiments.

%---------------------

\subsection{Pointwise Differentiation in the Partial Assembly Framework}
\label{sec_diff_pwise}

The singly-diagonally implicit Runge-Kutta (SDIRK) methods considered in this
work require the solution of a nonlinear system at every implicit stage.
Efficient Newton-based solution of these systems
relies on the availability of Jacobian information.
For high-order FE discretizations, however, computing Jacobians can be
challenging due to the complexity of the underlying numerical kernels and the
large number of operations involved in evaluating the discrete residual.
Furthermore, applying automatic differentiation as a black-box approach to the
entire discretization often requires intrusive code modifications, changes to
data structures and data types, and can introduce significant memory and
computational overhead while reducing performance portability \cite{MFEM2024}.
These difficulties are compounded by the need to avoid globally assembled
Jacobian matrices, whose construction and storage can become prohibitively
expensive for large-scale high-order simulations.

Instead, we leverage MFEM's finite element operator decomposition
\eqref{eq_decompose}, which makes explicit the separation between the
topological transfer operators ($P,G,B$) and the quadrature-point kernel $D$.
The transfer operators are linear and depend only on the mesh topology and the
finite element spaces; they do not depend on the solution,
physical coordinates, or other problem parameters \cite{MFEM2024}.
As a result, differentiation can be confined to the innermost
pointwise kernel evaluated at quadrature points, leading to
memory-efficient and AD-friendly derivative computations. In addition, these operations 
can be done in parallel via batched computation.

To illustrate, consider a nonlinear operator
written in partially assembled form as:
\begin{equation}
\label{eq_pa_nonlinear}
  \mathbf{F}(\mathbf{U}) = P^T G^T B^T \, D\!\left(B G P\,\mathbf{U}\right),
\end{equation}
where the operator $D$ acts independently at each quadrature point and may be
nonlinear in its inputs (e.g., values and/or gradients of the discrete fields).
Given a direction $\mathbf{dU}$, the chain rule yields the Jacobian action
\begin{equation}
\label{eq_pa_jac_action}
  \mathbf{J}(\mathbf{U})\,\mathbf{dU} = P^T G^T B^T
    \left[ \frac{\partial D}{\partial \mathbf{U}_q}(B G P\,\mathbf{U})
     \right] B G P\,\mathbf{dU},
\end{equation}
where $\mathbf{U}_q$ denotes the Q-vector variables associated with the T-vector
variables $\mathbf{U}$, i.e.\ the collection of quadrature point inputs for $D$.
Crucially, because $D$ is pointwise, its derivative
$\partial D/\partial \mathbf{U}_q$ is also pointwise: it is block-diagonal with
small dense blocks, one per quadrature point, and can therefore be evaluated
locally and in parallel.
Equations \eqref{eq_pa_nonlinear}-\eqref{eq_pa_jac_action} show that Jacobian
actions (and similarly transpose actions) inherit the same gather-apply-scatter
structure as the original operator \eqref{eq_pa_nonlinear},
with the only differentiated component being the quadrature-point kernel.

The same mechanism applies to derivatives with
respect to problem parameters and geometry.
In particular, for moving-mesh Lagrangian formulations, geometric factors such
as the element Jacobian and its determinant enter through quadrature-point
computations; derivatives with respect to nodal coordinates are therefore
again localized to per-element, per-quadrature-point operations.

For the Lagrangian hydrodynamics system considered here,
$\mathbf{U}=[\mathbf{x},\mathbf{v},\mathbf{e}]$, the dominant residual
evaluations consist of the partially assembled mass and force operators in
equation \eqref{eq_matrices}, whose quadrature-point kernels compute
equation-of-state, stress, and artificial viscosity contributions.
By expressing these kernels in MFEM's \dfem interface, the required pointwise
derivatives for Newton-Krylov solves are generated automatically and reused in
the same matrix-free or fully-assembled operator application path.

%---------------------

\subsection{AD with Enzyme}

Section~\ref{sec_diff_pwise} reduces the differentiation problem to computing
derivatives of the quadrature-point kernel $D$.
To obtain these derivatives efficiently without invasive rewrites of the full
discretization, we use Enzyme \cite{Enzyme2021}, an automatic differentiation
tool that operates on the LLVM intermediate representation (IR).
Because Enzyme works at the compiler IR level, it can differentiate code
generated from any language targeting LLVM, and it can leverage LLVM's existing
optimization pipeline to produce efficient derivative code for both CPU and GPU
backends.

In MFEM's \dfem interface, the quadrature-point kernel is implemented as a
compact pointwise function that maps the local values and gradients of the
discrete FE fields to residual contributions.
Enzyme is applied to these kernels to generate the corresponding
directional derivatives needed for Jacobian-vector actions.
The resulting derivative kernels are then composed with the unchanged transfer
operators ($P,G,B$) via the same partial assembly pipeline, yielding matrix-free
Jacobian applications with the same gather-apply-scatter structure as the
original operator and preserving quadrature-level parallelism.

%---------------------

\subsection{Application of \dfem in Laghos}

We demonstrate our approach by building on the open-source Laghos miniapp,
which is based on an explicit Lagrangian hydrodynamics implementation,
and reuse the same quadrature-point physics
kernels (e.g., equation-of-state, stress, and other pointwise calculations).
To enable differentiation in the partial assembly setting, the corresponding
discrete operators must be expressed in the \dfem interface by providing
wrappers for the full differentiable operator $\mathbf{F}(\mathbf{U})$.
The \dfem interface makes the data flow explicit by specifying which discrete
fields (and derived quantities such as values and gradients) enter the operator,
how they are transformed through the $P,G,B$ transfers, what outputs
are produced, and with respect to which variables derivatives are taken.
Internally, \dfem exploits this decomposition to apply the chain rule and
localize differentiation to the pointwise kernel $D$, while reusing the same
topological transfer operators in both residual and Jacobian applications.
This organization introduces some code overhead compared to the original
handwritten explicit operators, but it makes the differentiation targets
well-defined and allows Enzyme-generated derivatives to be integrated cleanly
into the partial assembly pipeline.

Importantly, this refactoring does not change the underlying spatial
discretization or the physical model: the \dfem operators call the same
pointwise functions used in the original code path, but through a uniform
interface that supports both residual evaluation and Jacobian actions.
As demonstrated by the Taylor-Green results in Section~\ref{sec_test_tg} and
Table~\ref{tab_TG_rates}, the reformulated explicit configuration (\dfem-E)
produces results equivalent to the original explicit method (ORIG-E) with only
modest performance overhead.
The same \dfem-based operators are then reused by the implicit solver to obtain
matrix-free Jacobian applications and, when needed, sparse matrix assembly for
preconditioning, without implementing new physics kernels from scratch.
One notable exception is the artificial viscosity model: for robust Newton-based
implicit solves we must employ a smoother viscosity formulation, as described in
Section~\ref{sec_diff_av}.

%---------------------

\subsection{Smoothed Artificial Viscosity}
\label{sec_diff_av}

Artificial viscosity is a critical component of robust shock capturing in
Lagrangian hydrodynamics. 
We write the total stress as $\sigma=-pI+\sigma_a$, where $\sigma_a$ is an
artificial stress used for shock capturing.
The artificial stress has the form
\begin{equation}
\label{eq_visc_stress}
  \sigma_a = \mu \, \varepsilon(v), \qquad
  \varepsilon(v)=\frac{1}{2}\left(\nabla v + (\nabla v)^T\right),
\end{equation}
with a viscosity coefficient $\mu$ that depends on the local flow state.

The artificial viscosity model used in Laghos is the tensor viscosity described in
Section~6 of~\cite{Dobrev2012}, which we denote here by the ``explicit''
viscosity coefficient $\mu_{\mathrm{exp}}$. 
In the original explicit formulation,
$\mu=\mu_{\mathrm{exp}}$ is defined using a directional compression measure based
on the most compressive eigenpair of $\varepsilon(v)$.
Specifically, letting $(\lambda_1,s_1)$ denote the smallest eigenvalue and its
eigenvector, the directional compression measure is
$\Delta v_{s_1}=s_1\cdot \varepsilon(v)\cdot s_1=\lambda_1$ and the viscosity
coefficient takes the form \cite[Section~6.2]{Dobrev2012}:
\begin{equation}
\label{eq_visc_mu_exp}
  \mu_{\mathrm{exp}} = \rho \left(q_2 \ell_{s_1}^2 |\Delta v_{s_1}| +
                                  q_1\,\psi_1\,\psi_0\,\ell_{s_1}\,c_s\right),
\end{equation}
where $\ell_{s_1}$ is a directional length scale and $\psi_1,\psi_0$ are
compression and vorticity switches.
While this design is effective for explicit shock capturing, it is problematic
for Newton-based implicit solves: eigenvalue crossings can switch the selected
eigenpair between Newton updates, which introduces non-smooth changes in both
the compression measure and the directional length scale.
Furthermore, the sharp compression switch and absolute value nonlinearity can
lead to large local derivative variations.
Combined with large implicit steps that move shocks across multiple elements,
these effects can degrade nonlinear convergence.
We have observed these challenges in our tests.

To obtain a smoother and more robust viscosity model for implicit time
integration, we replace the explicit coefficient $\mu_{\mathrm{exp}}$ by a new
coefficient $\mu_{\mathrm{imp}}$ that avoids eigenpair-dependent switching and
replaces non-differentiable pointwise operations with smooth approximations.
A similar implicit viscosity term is proposed in \cite{Korner2026}; here we
introduce a modified formulation aimed at improving smoothness and Newton
robustness.
The viscosity term retains the tensor form \eqref{eq_visc_stress}, with
$\mu=\mu_{\mathrm{imp}}$, which is defined by
\begin{equation}
	\label{eq_visc_mu_imp}
	\mu_{\mathrm{imp}}
	= \rho\,h\left(q_2\,|\delta_v|_{\delta_{\mathrm{scale}}}
	+ q_1\,c_s\,\psi(\delta_v)\right).
\end{equation}
In this formula, $h$ is an isotropic measure of the current element size,
$\delta_v$ is an isotropic compression measure,
$\psi$ is a smooth compression switch, and
$|\delta_v|_{\delta_{\mathrm{scale}}}$ is a smooth approximation of the absolute
value.
We set $q_1=0.5$ and $q_2=1$ by default, which yields shock sharpness
comparable to the original viscosity (see the discussion later in the section).

Let $J$ and $J_0$ denote the current and initial element Jacobians,
respectively, and let $h_0$ be a smoothed (e.g. a constant for close-to-uniform
meshes) characteristic physical length scale for the elements on the initial mesh.
We define an isotropic current length scale by volume scaling,
\begin{equation}
	\label{eq_visc_imp_h}
	h = h_0 \left(\frac{\det(J)}{\det(J_0)}\right)^{1/d},
\end{equation}
where $d$ is the spatial dimension. Using the physical velocity gradient
$\nabla v$, we then define the scalar compression measure
\begin{equation}
	\label{eq_visc_imp_dv}
	\delta_v = h \, \nabla \cdot v = h \, \mathrm{tr}(\nabla v),
\end{equation}
which has units of velocity and is isotropic by construction.

\begin{figure}[pos=htbp]
\hfill\includegraphics[width=0.45\textwidth]{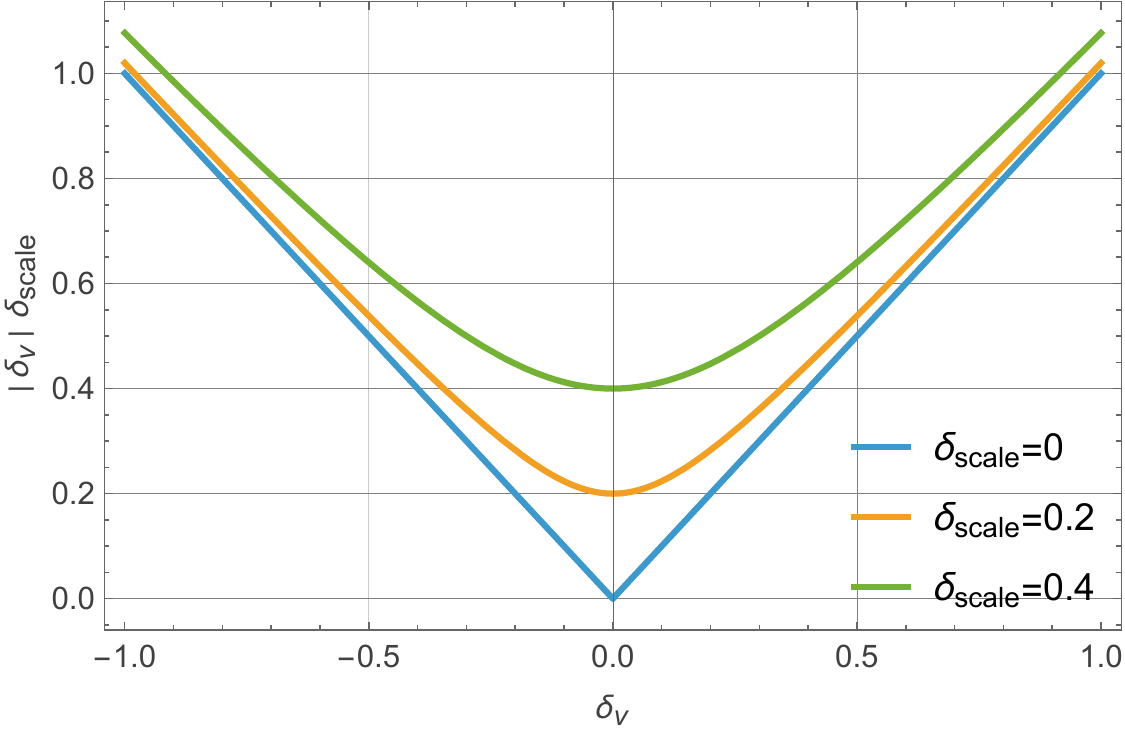}
\hfill\includegraphics[width=0.45\textwidth]{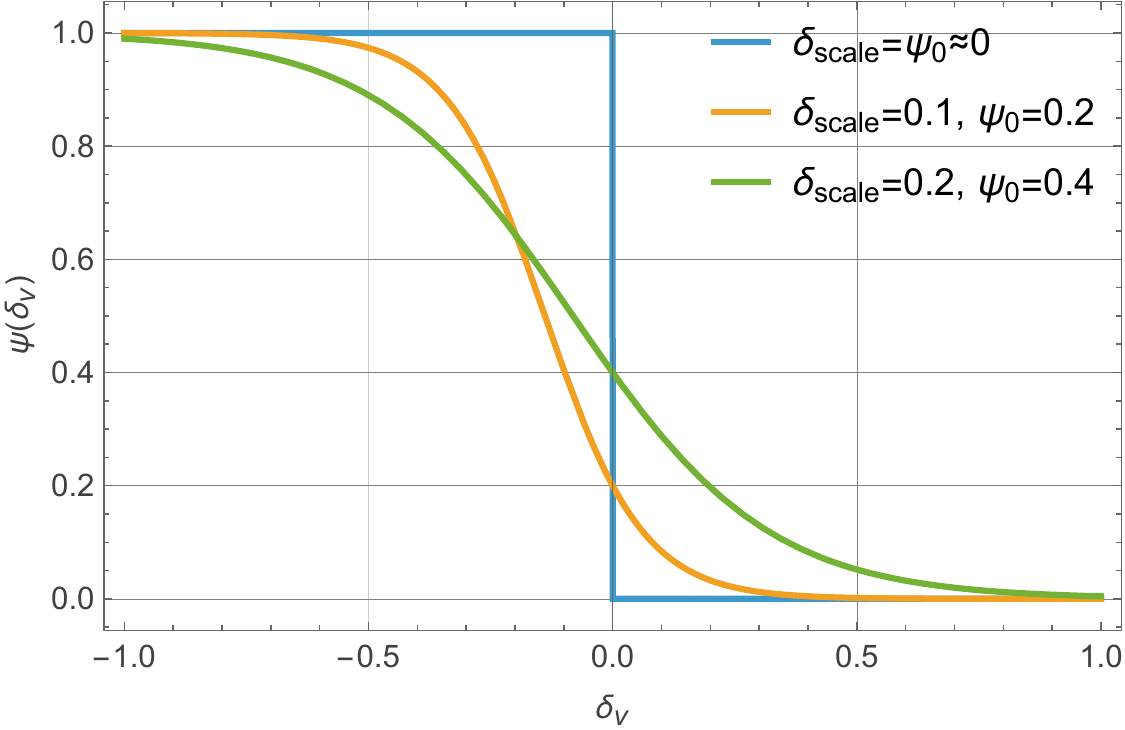}\hfill\hfill
\caption{Smoothed absolute value (left) and smoothed jump (right) functions.}
\label{fig_smooth_functions}
\end{figure}

The non-differentiable absolute value $|\delta_v|$ is replaced by the smooth
approximation
\begin{equation}
\label{eq_visc_imp_abs}
  |\delta_v|_{\delta_{\mathrm{scale}}} =
  \sqrt{\delta_v^2+\delta_{\mathrm{scale}}^2},
\end{equation}
which removes the kink at $\delta_v=0$. The parameter
$\delta_{\mathrm{scale}}$ controls the width of the smoothing region and
therefore the transition between the regularized and exact behavior; 
see Figure~\ref{fig_smooth_functions}.

To activate the viscosity smoothly under compression, we define a compression
switch $\psi\in(0,1)$ using a shifted hyperbolic tangent applied to $-\delta_v$,
so that compression corresponds to increasing argument:
\begin{equation}
\label{eq_visc_imp_switch}
 \psi(\delta_v) = \frac{1}{2} +
   \frac{1}{2} \tanh\!\left(\frac{-\delta_v-s}{2\,\delta_{\mathrm{scale}}}\right),
   \qquad s = 2 \delta_{\mathrm{scale}}
              \mathrm{atanh} (1 - 2 \psi_0).
\end{equation}
The parameter $\delta_{\mathrm{scale}}$ also determines the width of the
transition region between the inactive and active states of the switch.
The shift $s$ is chosen so that $\psi(0) = \psi_0$;
see Figure \ref{fig_smooth_functions}. In practice we use $\psi_0 = 0.5$
(numerical experiments with $h$-dependent values showed undesirable increase in
the number of Newton iterations in the nonlinear solves).
Both regularizations are controlled by the scaling
\begin{equation}
\label{eq_visc_imp_scale}
  \delta_{\mathrm{scale}} = 2 \frac{h}{h_{\mathrm{coarse}}}\,c_s,
\end{equation}
where $h_{\mathrm{coarse}}$ is a characteristic coarse mesh length scale
(such that $h \le h_{\mathrm{coarse}}$ in all cases) and
$c_s$ is the local speed of sound.
The parameter $\delta_{\mathrm{scale}}$ has units of velocity, matching the
units of the compression measure $\delta_v=h\nabla\cdot v$.
We choose it proportional to the local acoustic sound speed, $c_s$,
with the dimensionless factor $h/h_{\mathrm{coarse}}$ included so that the
smoothing and activation widths decrease under mesh refinement.
Thus, the regularization is large enough to provide a smooth nonlinear residual
on coarse meshes, while becoming sharper as the mesh is refined.

To keep $\delta_{\mathrm{scale}}$ in \eqref{eq_visc_imp_scale} strictly
positive when the specific internal energy vanishes, as it does in parts of the
initial condition for the Sedov test in Section~\ref{sec_test_sedov},
we evaluate the equation of state using the smooth positive part
\begin{equation}
\label{eq_energy_regularization}
  e_{+} = \frac{1}{2} \left( e + \sqrt{e^2+\varepsilon_e^2} \right),
  \qquad \varepsilon_e = 10^{-6},
\end{equation}
and define
\begin{equation}
  p=(\gamma-1)\rho e_{+}, \qquad
  c_s=\sqrt{\gamma(\gamma-1)e_{+}}.
\end{equation}
In particular, at $e=0$ we have $e_{+}=\varepsilon_e/2>0$, so both
$c_s$ and its derivative with respect to $e$ remain finite.
This regularization is imposed explicitly in the physical kernel and does
not rely on any special treatment of $\sqrt{0}$ by Enzyme.

\begin{figure}[pos=htbp]
\centering
\begin{tabular}{cc}
  \includegraphics[width=0.174\textwidth]{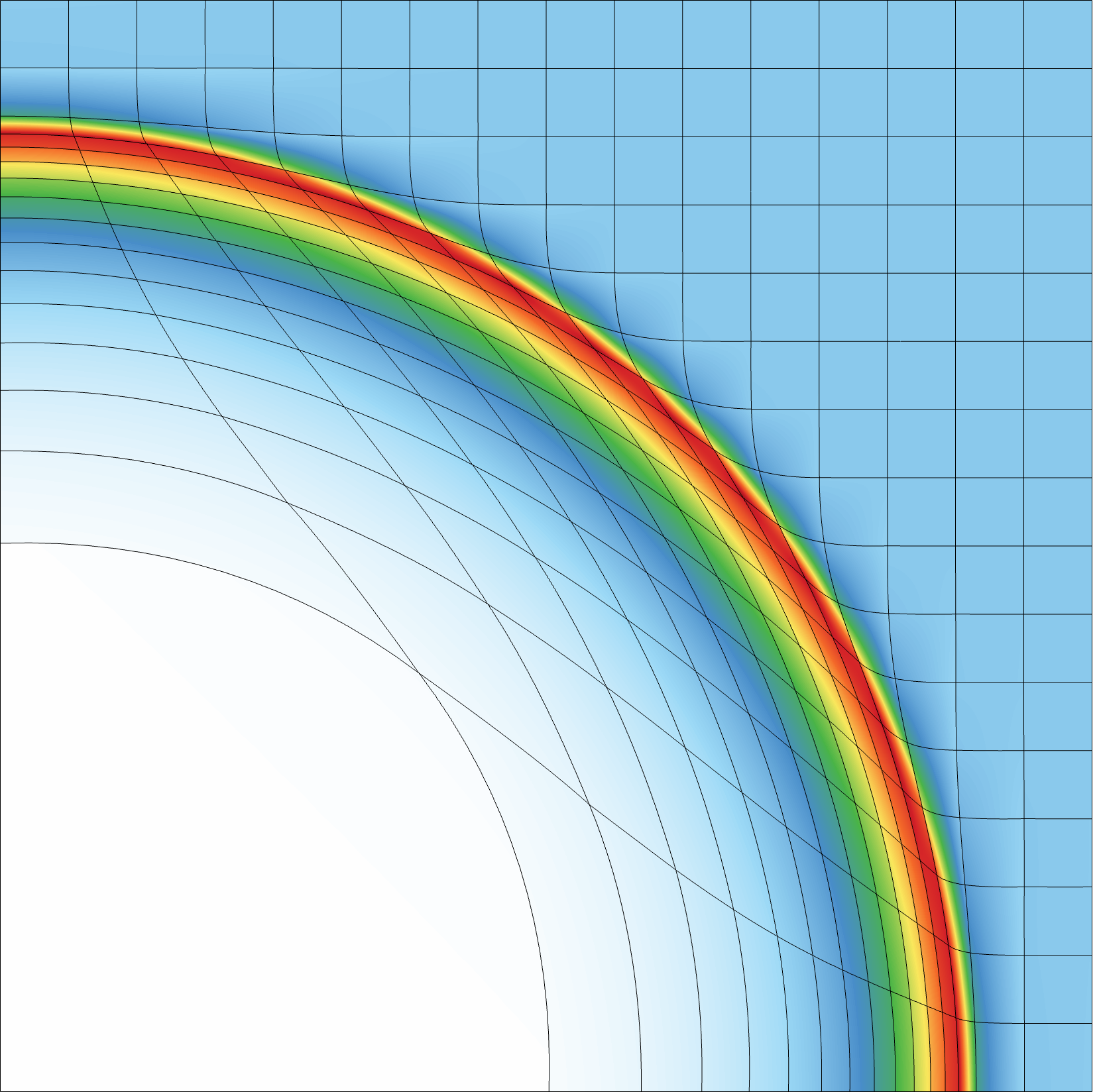} &
  \includegraphics[width=0.4\textwidth]{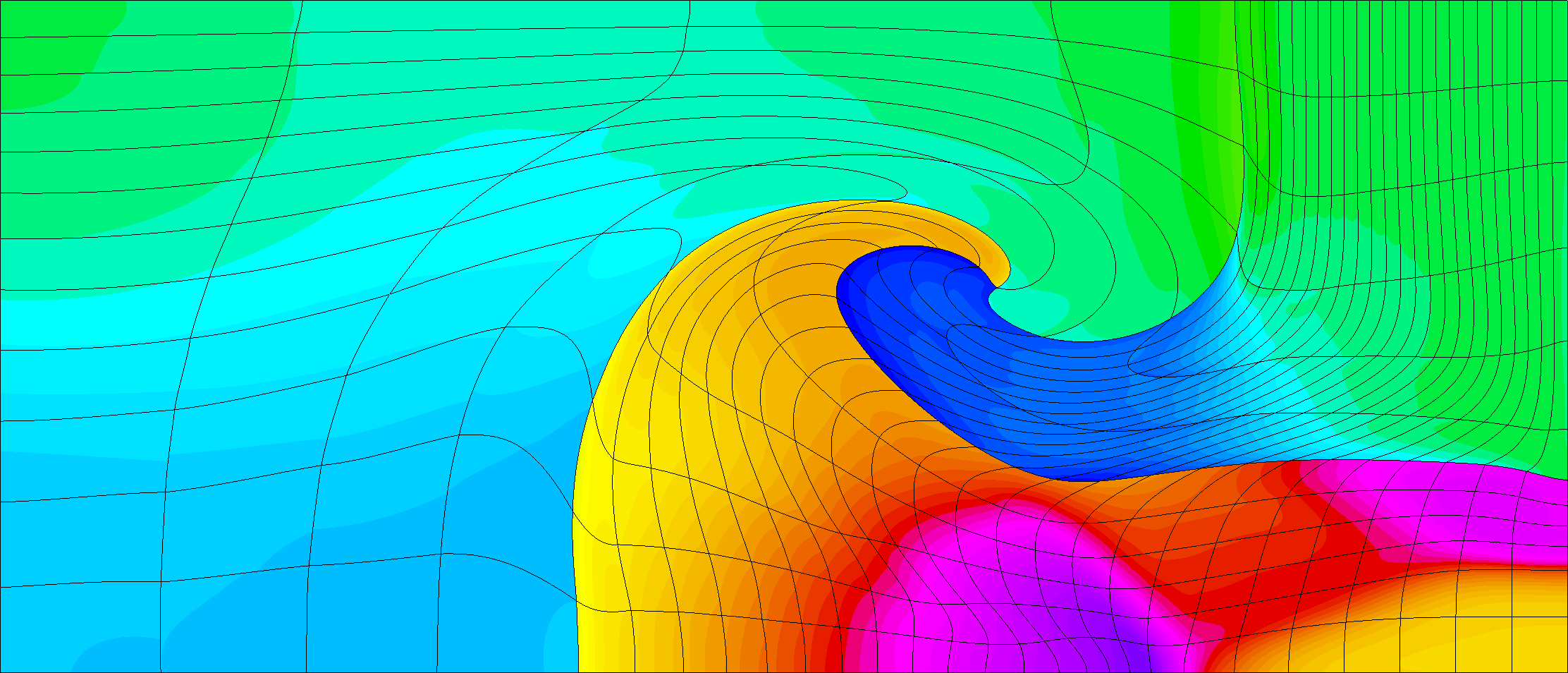} \\
  \includegraphics[width=0.174\textwidth]{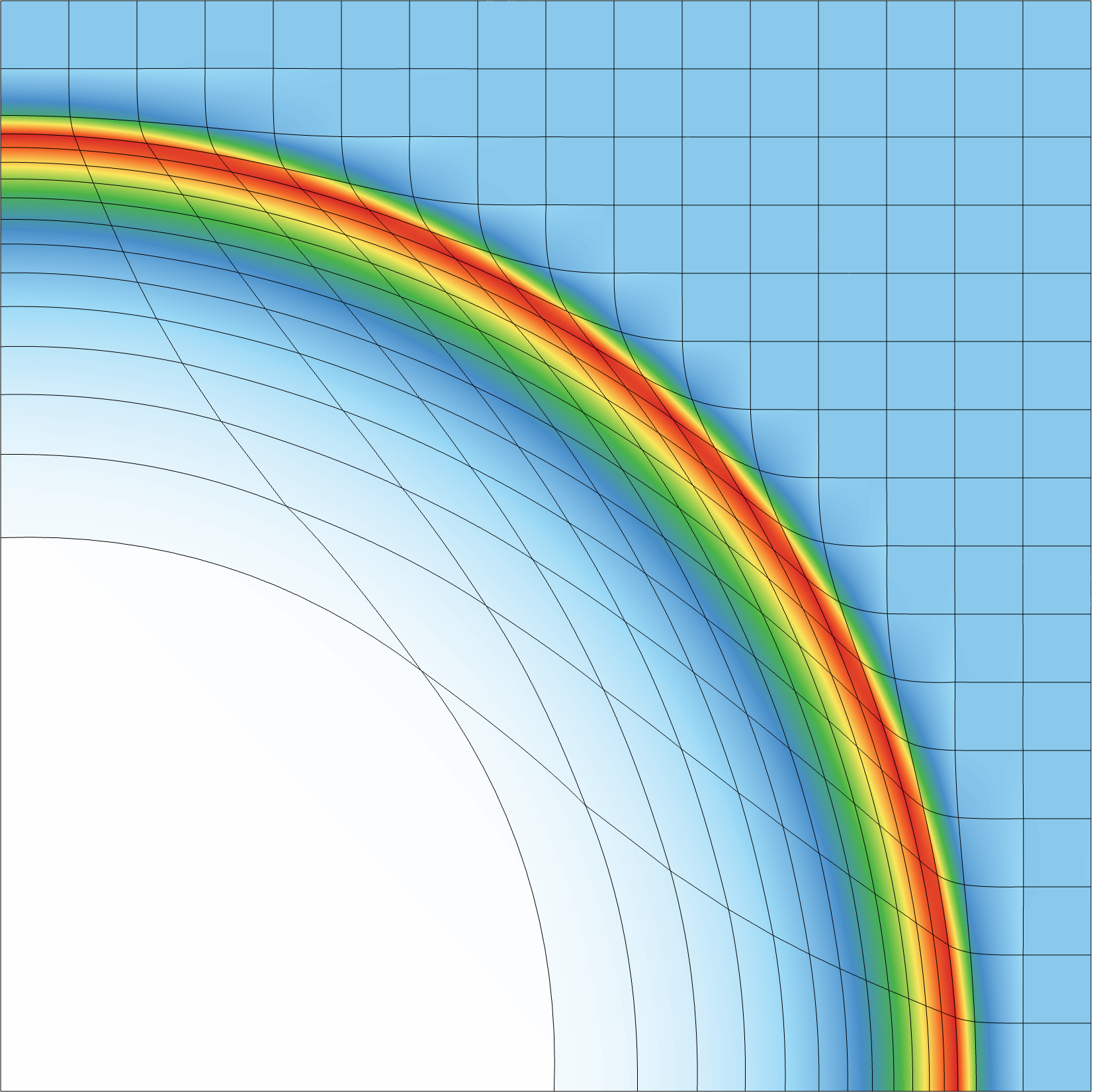} &
  \includegraphics[width=0.4\textwidth]{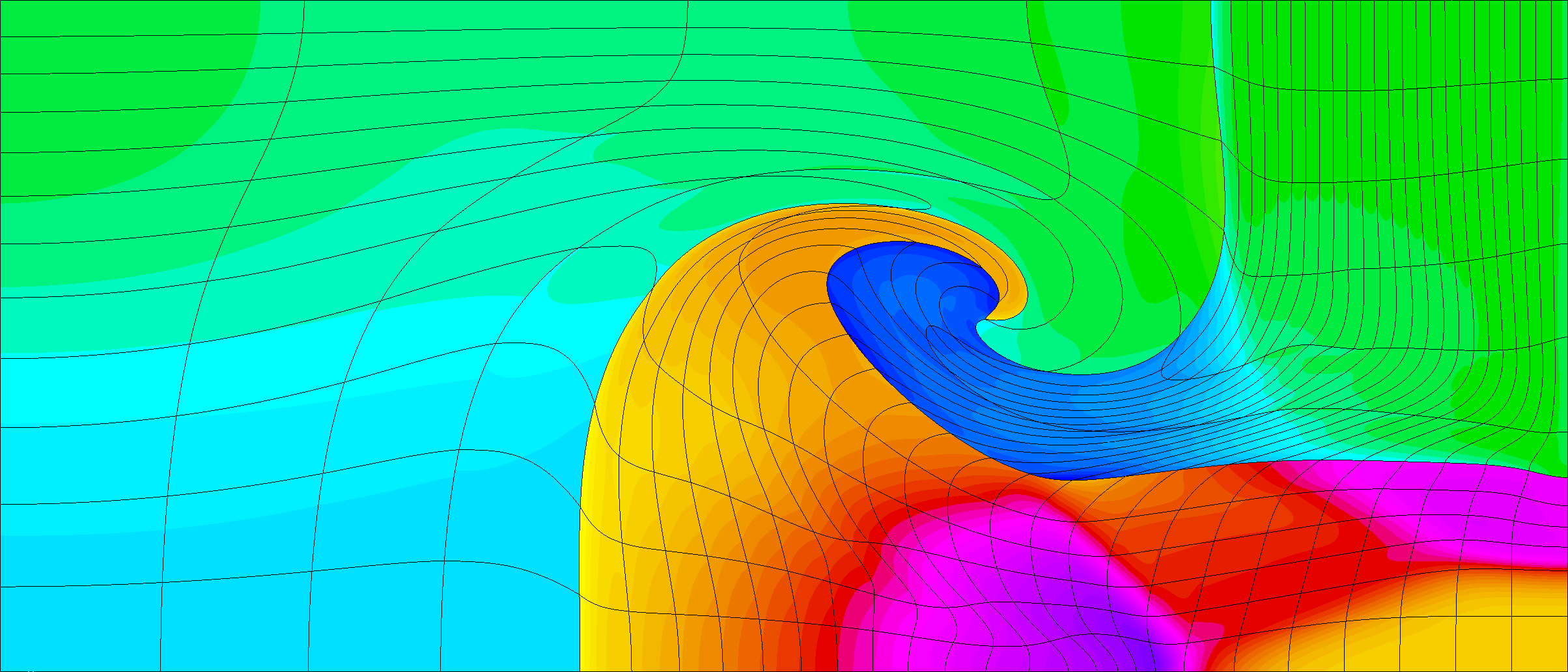}
\end{tabular}
\caption{Comparison of the density fields when using the original tensor
viscosity $\mu_{\mathrm{exp}}$ (top row) and the smoothed viscosity
$\mu_{\mathrm{imp}}$ (bottom row), with both formulations evaluated in explicit
time integration. Left: Q2Q1 Sedov Blast. Right: Q3Q2 Triple Point.}
\label{fig_visc_compare}
\end{figure}

The coefficients $q_1$ and $q_2$ in \eqref{eq_visc_mu_imp} were selected so
that, when both viscosity models are used in explicit time integration, the
smoothed coefficient $\mu_{\mathrm{imp}}$ approximately matches the shock
sharpness obtained with the original coefficient $\mu_{\mathrm{exp}}$ in
strongly compressive regions.
Figure \ref{fig_visc_compare} compares the two viscosity formulations on the
Sedov and Triple Point problems under this explicit-time-integration setting,
with the corresponding results plotted on the same scale.

These comparisons confirm that the two formulations produce similar shock
sharpness in the radially compressive Sedov problem and in the shock structures
in lower-right region of the Triple Point problem.
Away from these predominantly compressive regions, however,
the smoothed isotropic viscosity is less diffusive.
This is expected because $\mu_{\mathrm{imp}}$ is based on the trace-based measure
$\nabla\cdot v$, whereas $\mu_{\mathrm{exp}}$ detects directional compression
through the most compressive eigenpair of the symmetric velocity gradient.
In regions with strong shear and roll-up, compression in one direction can be
partially balanced by expansion in another, reducing the isotropic compression
measure even when directional compression remains significant.
This is evident in the Triple Point results, where the vorticity region
is diffused less by $\mu_{\mathrm{imp}}$. Note that this deficiency in the
results with $\mu_{\mathrm{exp}}$ can be eliminated by enabling the vorticity
switch $\psi_0$ in \eqref{eq_visc_mu_exp} (see Section 6.2 in \cite{Dobrev2012})
which was set to $1$ in the numerical tests presented here.

Finally, we note that both the explicit formulation of \cite{Dobrev2012} and 
the smoothed version described here are missing hyper-viscosity limiting as 
described in \cite{Maldonado2020}. Limiting techniques like this, which blend
first-order  dissipation for shocks with high-order dissipation for smooth regions, 
are imperative for preventing excess numerical dissipation throughout a calculation and 
for preserving high-order convergence in smooth flows.
Developing differentiable approaches for hyper-viscosity
limiting is a topic of future research.

%---------------------

\subsection{Solving the Nonlinear System}

Implicit SDIRK time integration requires the solution of a
nonlinear system at each implicit stage.
In what follows, $\mathbf{k}, \mathbf{u}, \mathbf{w}$ are block vectors with
components for position, velocity, and internal energy.
In MFEM's interface, each stage solve is
posed in terms of a stage increment $\mathbf{k}$ satisfying
\begin{equation}
		\label{eq_nonlinear_system}
		\mathbf{R}(\mathbf{k};\tau,\mathbf{u}) = 0,
\end{equation}
where $\tau$ is the stage scaling (proportional to the time step),
$\mathbf{u}$ is the current stage base state, and
$\mathbf{R}$ is formed by evaluating the hydrodynamics operators at the
stage state $\mathbf{w}=\mathbf{u}+\tau\mathbf{k}$.

We solve \eqref{eq_nonlinear_system} using an inexact Newton--Krylov method.
Given an iterate $\mathbf{k}^m$, we compute the residual
$\mathbf{r}^m=\mathbf{R}(\mathbf{k}^m)$ and solve the linearized system
\begin{equation}
\label{eq_newton_linear}
	\mathbf{J}(\mathbf{k}^m)\,\Delta \mathbf{k}^m = -\mathbf{r}^m,
\end{equation}
where $\mathbf{J}=\partial \mathbf{R}/\partial \mathbf{k}$.
The Jacobian is applied in a matrix-free manner using the derivative operators
described in Section~\ref{sec_diff_pwise}, and the linear system
\eqref{eq_newton_linear} is solved with restarted GMRES.
To prevent mesh tangling, we augment Newton with a backtracking line search that
rejects trial updates that would produce negative element Jacobian determinants.
The Newton iteration uses a default relative tolerance of $10^{-5}$, while the
GMRES solve uses a relative tolerance of $10^{-12}$. Both solvers are limited
to a maximum of 10 iterations.

Efficient preconditioning for \eqref{eq_newton_linear} is essential for
scalable implicit performance.
By default, we use a block-diagonal AMG preconditioner, which was found to give
the best parallel scalability in our experiments and is therefore the default
in the implementation.
The preconditioner is built by assembling the diagonal Jacobian blocks
corresponding to the velocity and energy components (combined with the
associated mass-matrix terms arising in the implicit stage residual) into
sparse matrices, and applying Hypre BoomerAMG \cite{hypre}
to each block independently. This yields an operator of the form:
\begin{equation}
\label{eq_blockdiag_prec}
  \mathbf{P}^{-1} \approx
  \mathrm{diag}\!\left(I,\;
  \mathrm{AMG}\!\left(\mathbf{M}_\mathcal{V} +
                      \tau\,\partial \mathbf{F}/\partial \mathbf{v}\right),\;
  \mathrm{AMG}\!\left(\mathbf{M}_\mathcal{E} -
                      \tau\,\partial \mathbf{F}^\top/\partial \mathbf{e}\right)\right),
\end{equation}
where $\mathbf{F}$ denotes the discrete force operator from \eqref{eq_matrices}.
We also support an optional PETSc SNES solver \cite{PETSC2026} path
based on an assembled Jacobian and sparse direct linear solve.
To reduce assembly costs, Jacobian updates can be lagged across nonlinear
iterations; for example, all implicit stages of an SDIRK step may reuse the
same assembled Jacobian during their nonlinear solves.
In our serial experiments, this configuration consistently delivered the best
performance, and its use is explicitly identified in the reported results.

%-------------------------------------------------

\section{Numerical Tests}
\label{sec_tests}

In this section we present numerical results on standard benchmark problems to
demonstrate the behavior of the proposed implicit method.
We first verify high-order convergence on the smooth Taylor-Green vortex
(in the absence  of artificial viscosity).
We then demonstrate correct shock behavior on the Sedov blast problem using the
new smoothed artificial viscosity.
Finally, we show improved accuracy-per-time-to-solution on the Triple Point
problem, where explicit stability restrictions become increasingly severe at
later times, i.e.\ the explicit time steps become very small.

All simulations are performed using the MFEM \cite{MFEM2020} library and utilize
its \dfem interface, which employs Enzyme for automatic differentiation-based
construction of the Jacobian of the discrete nonlinear residual operator.
The resulting nonlinear systems are solved with a standard Newton method
augmented by a line search strategy to prevent mesh inversion.
The Jacobian operator is applied in a matrix-free manner throughout.
Unless stated otherwise, the linear systems arising in the Newton iterations
are preconditioned with a block-diagonal AMG method, which was found to
provide the best parallel performance in our experiments.

\paragraph{Implicit time step control.}
For the implicit calculations, the specified CFL number is used only to multiply
the initial time step, following the explicit time step estimate used in
\cite{Dobrev2012}. To emphasize that this CFL number is used only for the first
time step, we will refer to it as first-step-CFL (FS-CFL).
After the first step, time step control is based on the
observed Newton convergence behavior. If the previous step required at most one
Newton iteration, the time step is increased by $5\%$. If it required two or
three Newton iterations, the time step is left unchanged. If four or more Newton
iterations were required, the step is rejected and repeated with $85\%$ of
the previous time step.

%---------------------

\subsection{2D Taylor-Green}
\label{sec_test_tg}

The purpose of this example is to verify that the implicit method retains
high-order convergence for a smooth problem, even with large time steps.
Additionally, it demonstrates that for a given resolution, the implicit method
can outperform the explicit one in total runtime by requiring fewer time steps.
Since the artificial viscosities we consider here are missing a limiter, we
explicitly set this term to zero in this test. 

\begin{figure}[pos=htbp]
\centerline{%
\begin{minipage}[c]{0.3\textwidth}%
\centering
\includegraphics[width=\linewidth]{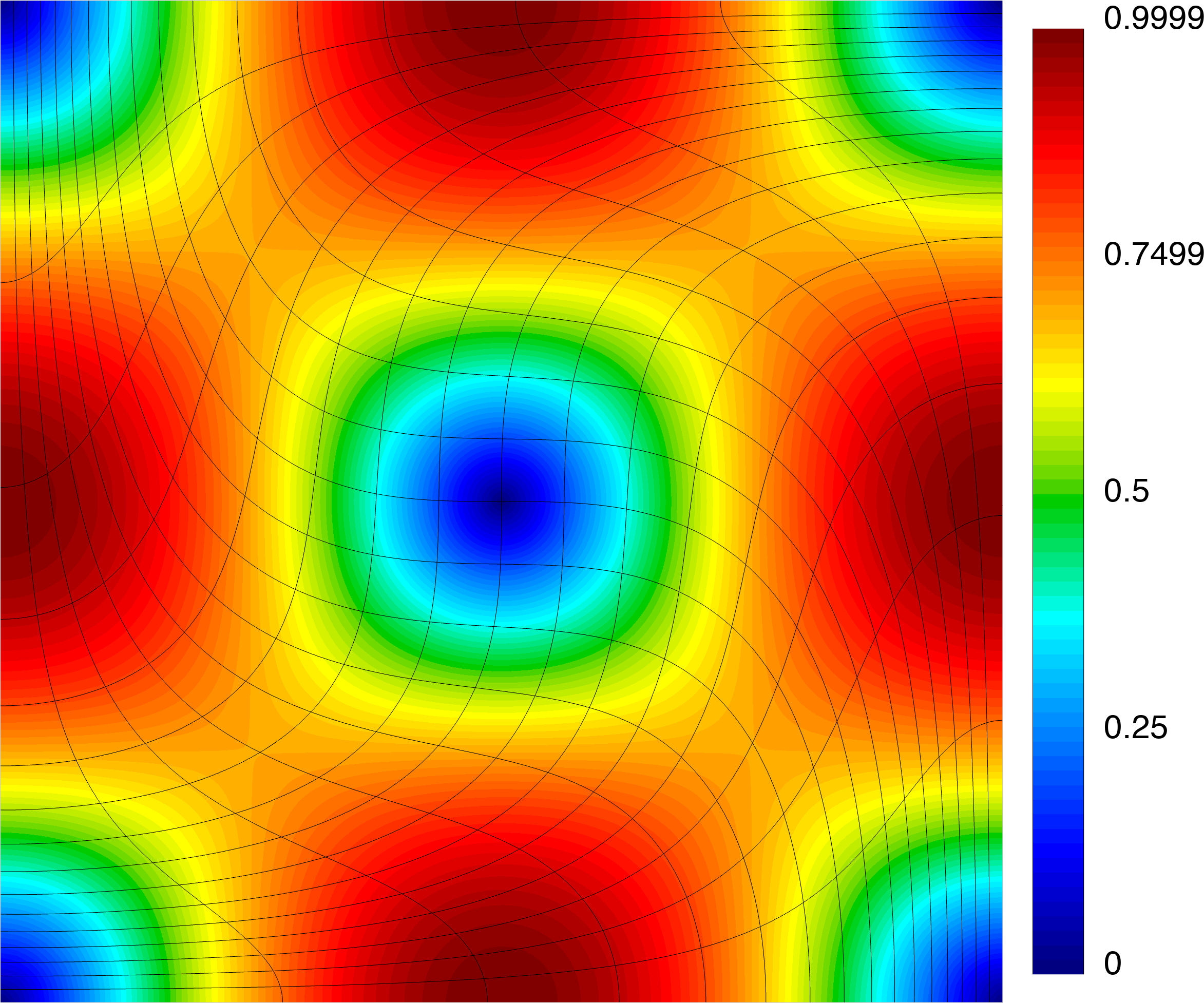}%
\end{minipage}%
\hspace{0.6em}%
\begin{minipage}[c]{0.295\textwidth}%
\centering
\includegraphics[width=\linewidth]{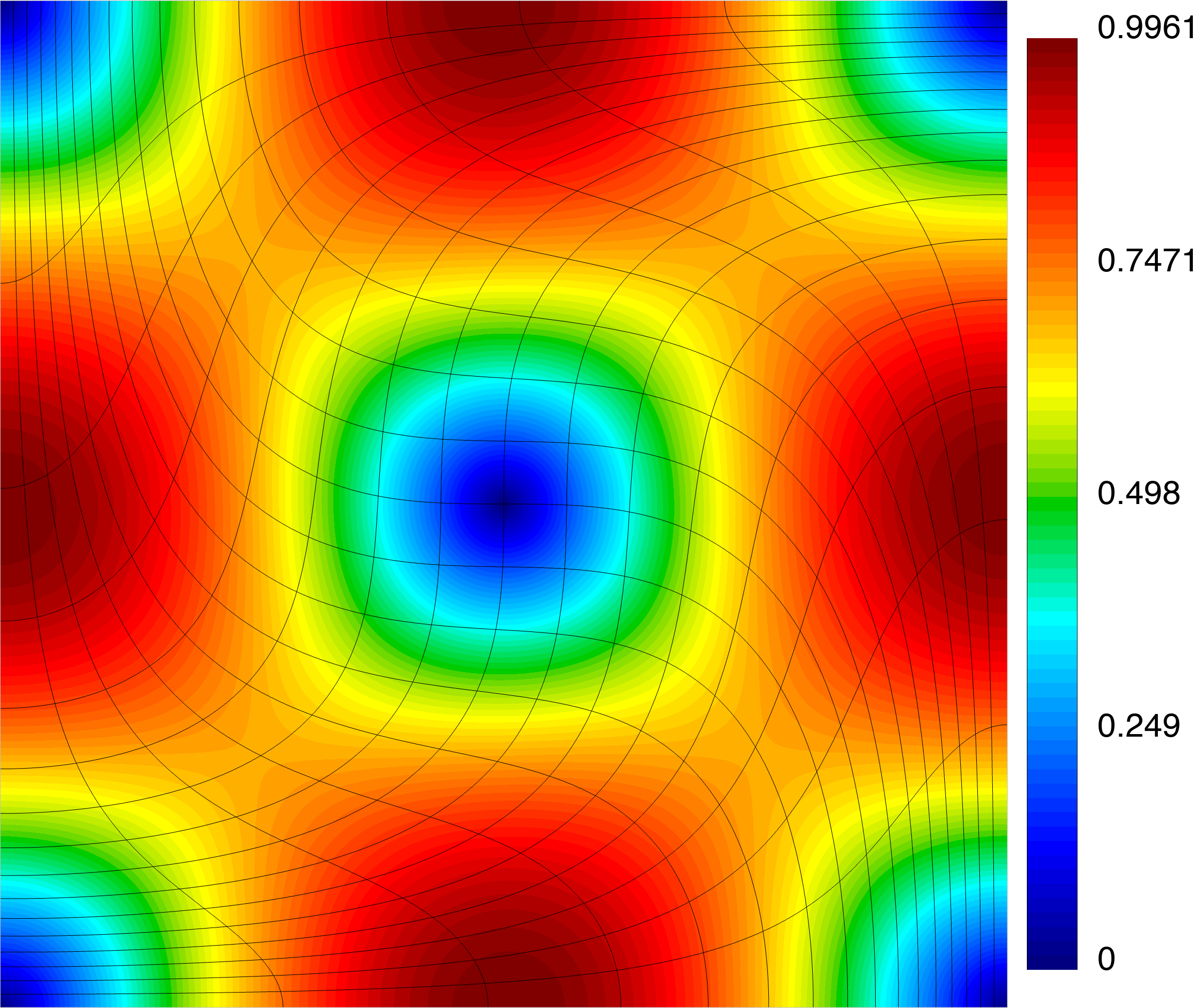}%
\end{minipage}%
\hspace{0.6em}%
\begin{minipage}[c]{0.32\textwidth}%
\centering
\includegraphics[width=\linewidth]{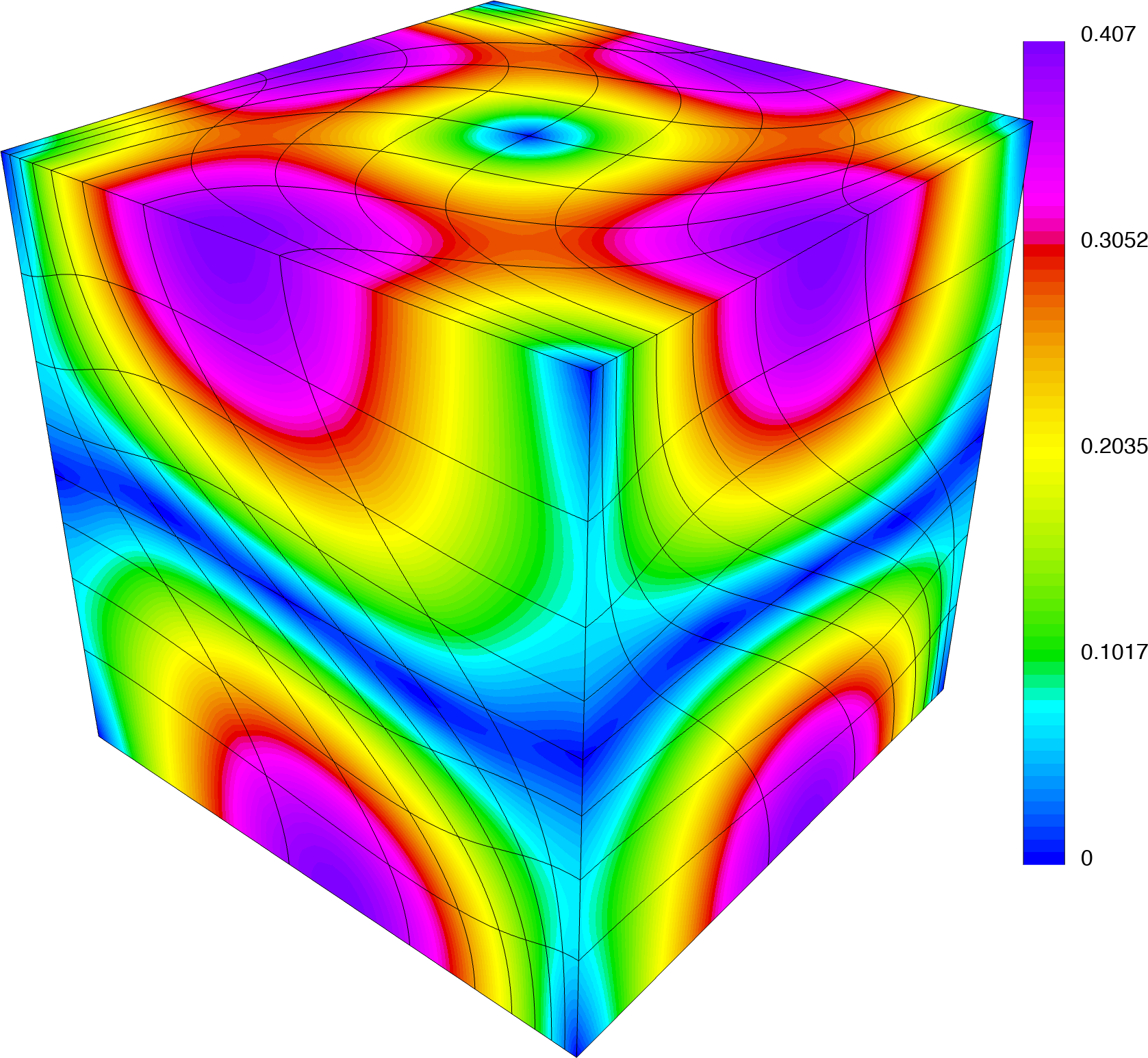}%
\end{minipage}%
}
\caption{Velocity magnitudes for a Q3Q2 spatial discretization of the
         Taylor-Green problem.
         Left: explicit calculation, 202 time steps.
         Middle: implicit calculation, 5 time steps.
         Right: 3D implicit calculation to time 0.75, 53 time steps.}
\label{fig_TG_v}
\end{figure}

The 2D Taylor-Green vortex problem is a standard benchmark used in many studies
of hydrodynamics methods to verify convergence order; its setup is described in
Section~8.1 of \cite{Dobrev2012}. Velocity magnitude and mesh position at
time 0.5 are shown in Figure~\ref{fig_TG_v}.
We compare three main execution configurations:
\begin{itemize}
\item \textit{ORIG-E}: explicit time integration corresponding to
      the original explicit method as presented in \cite{Dobrev2012}.
\item \textit{\dfem-E}: explicit time integration through the reformulated
      version of the original explicit method, with operators written in a form
      suitable for implicit computations.
\item \textit{\dfem-I}: the implicit time integration developed in this work.
\end{itemize}
For explicit time stepping, we use the following three methods: RK2: the
RK2-average scheme defined in Section 7.1 of \cite{Dobrev2012}; RK3: the
strong stability preserving RK3 method, implemented in MFEM's class
\code{RK3SSPSolver}; RK4: the standard RK4 method, implemented in MFEM's
class \code{RK4Solver}. For implicit time stepping we use the three methods we
denote here as SD2/SD3/SD4: these are the order 2/3/4 SDIRK methods implemented
in MFEM's classes \code{ImplicitMidpointSolver}, \code{SDIRK33Solver}, and
\code{SDIRK34Solver}, respectively.

The results are summarized in Table~\ref{tab_TG_rates}.
The explicit configurations are run with a default CFL number of 0.5,
while the \dfem-I method is tested with FS-CFL values of 0.5, 2, and 8.
In the explicit cases, a specialized time step control is applied to prevent
potential instabilities, as described in Section~7.3 of \cite{Dobrev2012}.
For \dfem-I, the time step is instead controlled by the Newton-based procedure
described above. Since the implicit formulation is not subject to the same
explicit stability restriction, this procedure can maintain or increase the time
step when the nonlinear solves remain inexpensive, whereas the explicit method
may need to repeat steps with reduced time step sizes. As a result, even with
the same FS-CFL value of $0.5$, \dfem-I requires significantly fewer time
steps than the explicit runs.

\begin{figure}[pos=htbp]
\begin{center}
      \begin{tabular}{@{}c@{\hspace{0.02\textwidth}}c@{\hspace{0.02\textwidth}}c@{}}
      \includegraphics[width=0.32\textwidth]{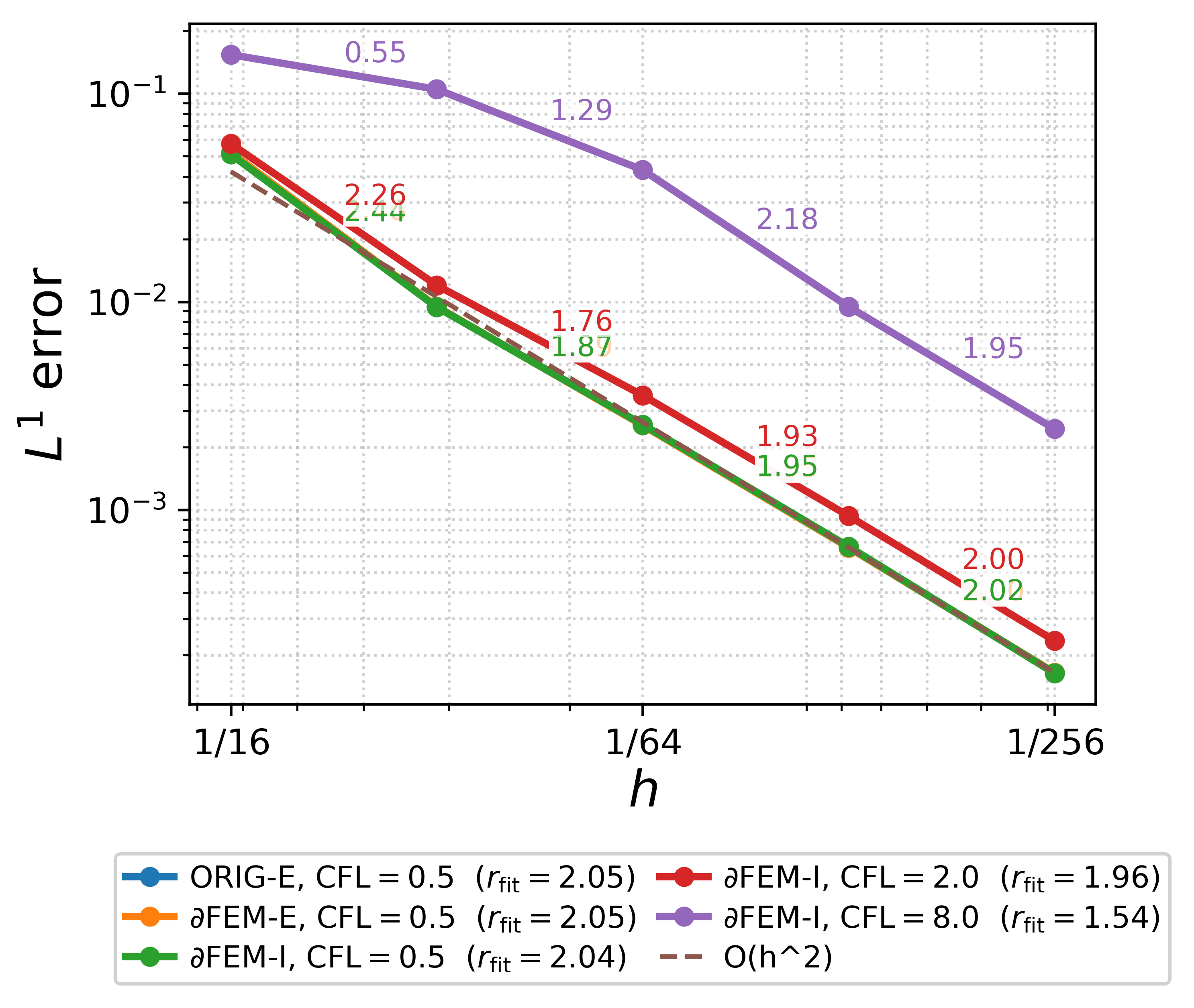} &
      \includegraphics[width=0.32\textwidth]{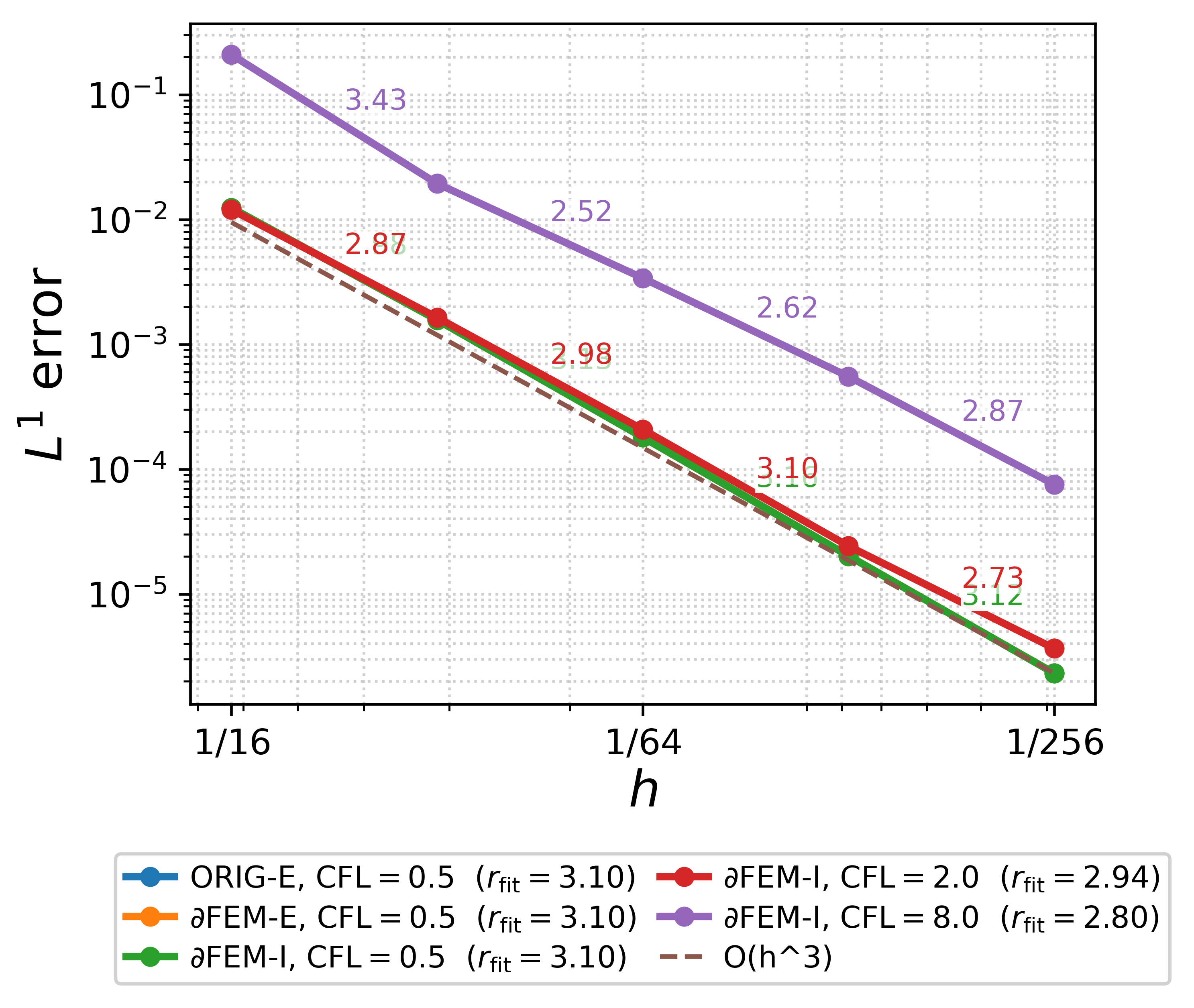} &
      \includegraphics[width=0.32\textwidth]{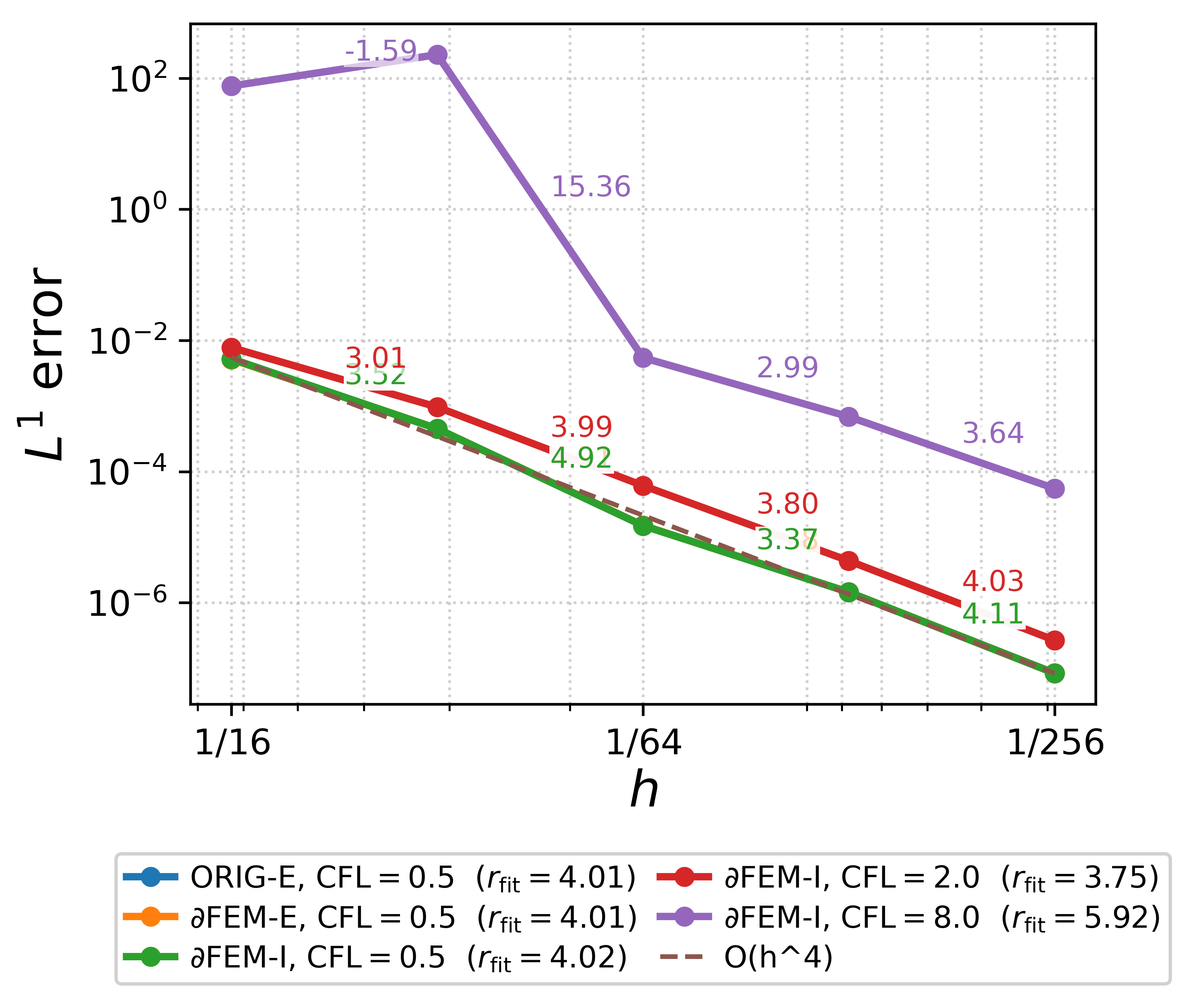} \\
      \textrm{(a) Q2Q1 + RK2/SD2} & \textrm{(b) Q3Q2 + RK3/SD3} &
      \textrm{(c) Q4Q3 + RK4/SD4} \\
\end{tabular}
\end{center}
\caption{Convergence rates for explicit and implicit time discretizations
         applied to the 2D Taylor-Green problem.
         Panels (a)--(c) show the Q2Q1, Q3Q2, and Q4Q3 discretizations, respectively.
         The explicit runs use $\mathrm{CFL}=0.5$, while the $\partial$FEM-I
         curves show results for $\mathrm{FS-CFL}=0.5$, $2.0$, and $8.0$.
         Numbers along the curves denote point-to-point convergence rates,
         and $r_{\mathrm{fit}}$ in the legend denotes
         the fitted log-log convergence rate.}
\label{fig_TG_rates}
\end{figure}

Table~\ref{tab_TG_rates} reports the velocity errors, convergence rates,
number of time steps, and wall-clock times for each run,
all executed on a single MPI task using the same standard laptop, to time 0.5.
These rates are also visualized in Figure~\ref{fig_TG_rates}.
The linear solve in all these tests uses the PETSc SNES solver
based on an assembled Jacobian and sparse direct linear solve.
The following observations can be drawn from Table~\ref{tab_TG_rates}:
\begin{enumerate}
\item When executed in explicit mode, the reformulated method (\dfem-E)
      produces results equivalent to the original method (ORIG-E), with some
      performance overhead.
      This overhead could be reduced through further implementation optimization.
\item The implicit runs (\dfem-I) achieve the expected high-order convergence.
      For example, fourth-order overall accuracy is demonstrated using a
      Q4Q3 spatial discretization.
      Even at large time steps (e.g., FS-CFL = 8), the observed convergence rates
      remain largely consistent with theoretical expectations.
\item While temporal errors increase at higher FS-CFL values, the implicit method
      remains robust at these larger time steps. In the explicit methods,
      stability is maintained by rejecting steps when necessary and repeating
      them with reduced time step sizes.
\item The runtime of the implicit method becomes comparable to the explicit
      one at around FS-CFL = 2. At FS-CFL = 8, implicit runs are consistently 3--4
      times faster. Further speedups are achievable with improved implementation
      and more efficient nonlinear solvers.
\end{enumerate}

\begin{table}[pos=htbp]
\begin{center}
  \resizebox{\columnwidth}{!} {
  \begin{tabular}{c | c c c c | c c c c | c c c c | c c c c | c c c c}
  \hline
        & \multicolumn{4}{c|}{ORIG-E CFL=0.5}  & \multicolumn{4}{c|}{\dfem-E CFL=0.5}  & \multicolumn{4}{c|}{\dfem-I FS-CFL=0.5}  & \multicolumn{4}{c|}{\dfem-I FS-CFL=2.0}  & \multicolumn{4}{c }{\dfem-I FS-CFL=8.0} \\
  \hline
        & \multicolumn{4}{c|}{Q2Q1 + RK2}      & \multicolumn{4}{c|}{Q2Q1 + RK2}      & \multicolumn{4}{c|}{Q2Q1 + SD2}      & \multicolumn{4}{c|}{Q2Q1 + SD2}      & \multicolumn{4}{c}{Q2Q1 + SD2}      \\
  \hline
  $h$   & $L^1$ error & rate & steps & time(s) & $L^1$ error & rate & steps & time(s) & $L^1$ error & rate & steps & time(s) & $L^1$ error & rate & steps & time(s) & $L^1$ error & rate & steps & time(s)\\
  \hline
  1/16  & 5.22E-02    & -    & 35    & 0.1     & 5.22E-02    & -    & 35    & 0.2     & 5.11E-02    & -    & 13    & 0.2     & 5.75E-02    & -    & 4     & 0.1     & 1.54E-01    & -    & 1     & 0.1    \\
  1/32  & 9.47E-03    & 2.46 & 68    & 0.3     & 9.47E-03    & 2.46 & 68    & 0.4     & 9.42E-03    & 2.44 & 26    & 0.3     & 1.20E-02    & 2.26 & 7     & 0.2     & 1.05E-01    & 0.55 & 2     & 0.1    \\
  1/64  & 2.55E-03    & 1.89 & 135   & 1.3     & 2.55E-03    & 1.89 & 135   & 1.9     & 2.57E-03    & 1.87 & 51    & 1.4     & 3.55E-03    & 1.76 & 12    & 0.6     & 4.30E-02    & 1.29 & 4     & 0.4    \\
  1/128 & 6.60E-04    & 1.95 & 268   & 8       & 6.60E-04    & 1.95 & 268   & 13      & 6.66E-04    & 1.95 & 102   & 12      & 9.35E-04    & 1.93 & 26    & 3.6     & 9.46E-03    & 2.18 & 7     & 2.1    \\
  1/256 & 1.65E-04    & 2.00 & 535   & 57      & 1.65E-04    & 2.00 & 535   & 92      & 1.64E-04    & 2.02 & 203   & 126     & 2.35E-04    & 2.00 & 51    & 34      & 2.45E-03    & 1.95 & 13    & 16     \\
  \hline
        & \multicolumn{4}{c|}{Q3Q2 + RK3}      & \multicolumn{4}{c|}{Q3Q2 + RK3}      & \multicolumn{4}{c|}{Q3Q2 + SD3}      & \multicolumn{4}{c|}{Q3Q2 + SD3}      & \multicolumn{4}{c}{Q3Q2 + SD3}      \\
  \hline
  $h$   & $L^1$ error & rate & steps & time(s) & $L^1$ error & rate & steps & time(s) & $L^1$ error & rate & steps & time(s) & $L^1$ error & rate & steps & time(s) & $L^1$ error & rate & steps & time(s)\\
  \hline
  1/16  & 1.24E-02    & -    & 51    & 0.2     & 1.24E-02    & -    & 51    & 0.2     & 1.24E-02    & -    & 20    & 0.3     & 1.20E-02    & -    & 5     & 0.2     & 2.09E-01    & -    & 2     & 0.1    \\
  1/32  & 1.57E-03    & 2.98 & 101   & 0.6     & 1.57E-03    & 2.98 & 101   & 0.9     & 1.57E-03    & 2.98 & 38    & 1.3     & 1.64E-03    & 2.87 & 10    & 0.5     & 1.94E-02    & 3.43 & 3     & 0.3    \\
  1/64  & 1.80E-04    & 3.13 & 202   & 3.8     & 1.80E-04    & 3.13 & 202   & 5.7     & 1.80E-04    & 3.13 & 76    & 11      & 2.08E-04    & 2.98 & 19    & 3.0     & 3.39E-03    & 2.52 & 5     & 1.6    \\
  1/128 & 2.02E-05    & 3.16 & 402   & 26      & 2.02E-05    & 3.16 & 402   & 40      & 2.02E-05    & 3.16 & 152   & 106     & 2.42E-05    & 3.10 & 38    & 26      & 5.52E-04    & 2.62 & 10    & 7.7    \\
  1/256 & 2.32E-06    & 3.12 & 802   & 200     & 2.32E-06    & 3.12 & 802   & 292     & 2.32E-06    & 3.12 & 304   & 1150    & 3.66E-06    & 2.73 & 76    & 288     & 7.54E-05    & 2.87 & 19    & 71     \\
  \hline
        & \multicolumn{4}{c|}{Q4Q3 + RK4}      & \multicolumn{4}{c|}{Q4Q3 + RK4}      & \multicolumn{4}{c|}{Q4Q3 + SD4}      & \multicolumn{4}{c|}{Q4Q3 + SD4}      & \multicolumn{4}{c}{Q4Q3 + SD4}      \\
  \hline
  $h$   & $L^1$ error & rate & steps & time(s) & $L^1$ error & rate & steps & time(s) & $L^1$ error & rate & steps & time(s) & $L^1$ error & rate & steps & time(s) & $L^1$ error & rate & steps & time(s)\\
  \hline
  1/16  & 5.16E-03    & -    & 68    & 0.3     & 5.16E-03    & -    & 68    & 0.7     & 5.25E-03    & -    & 26    & 0.7     & 7.77E-03    & -    & 7     & 0.4     & 7.62E+01    & -    & 2     & 0.2    \\
  1/32  & 4.50E-04    & 3.52 & 135   & 1.5     & 4.50E-04    & 3.52 & 135   & 1.9     & 4.50E-04    & 3.52 & 51    & 4.3     & 9.63E-04    & 3.01 & 13    & 1.4     & 2.29E+02    & -1.59& 4     & 1.0    \\
  1/64  & 1.50E-05    & 4.91 & 268   & 10      & 1.50E-05    & 4.91 & 268   & 12      & 1.49E-05    & 4.92 & 102   & 37      & 6.08E-05    & 3.99 & 26    & 9.5     & 5.45E-03    & 15.36& 7     & 5.1    \\
  1/128 & 1.44E-06    & 3.38 & 535   & 76      & 1.44E-06    & 3.38 & 535   & 91      & 1.44E-06    & 3.37 & 203   & 360     & 4.34E-06    & 3.80 & 51    & 87      & 6.86E-04    & 2.99 & 13    & 28     \\
  1/256 & 8.36E-08    & 4.11 & 1069  & 578     & 8.36E-08    & 4.11 & 1069  & 720     & 8.36E-08    & 4.11 & 405   & 3613    & 2.66E-07    & 4.03 & 102   & 906     & 5.50E-05    & 3.64 & 26    & 242    \\
  \hline
  \end{tabular} }
 \end{center}
 \caption{Convergence rates and execution times for explicit and implicit time
          discretizations applied to the 2D Taylor-Green problem.}
 \label{tab_TG_rates}
 \end{table}

 To further demonstrate that the implicit method extends naturally to three
 dimensions and behaves as expected, in Figure~\ref{fig_TG_v} we include
 a 3D result at time $t=0.75$, where the artificial viscosity is active.
 The Q3Q2 simulation was performed with a FS-CFL number of 8.0 and
 completed in only 40 time steps.

%---------------------

\subsection{2D Sedov Blast}
\label{sec_test_sedov}

Next we consider the 2D Sedov blast problem, a standard
verification case for strong-shock hydrodynamics.
The purpose of this test is to assess whether the implicit method
accurately resolves the outward-moving shock under large time steps,
preserves radial symmetry, and maintains physically reasonable mesh deformation.
It also serves to demonstrate the robustness of the new smoothed viscosity model in the
presence of steep gradients and discontinuities.

The computational domain is $[0,1] \times [0,1]$ with wall boundary conditions on all sides.
We consider a single material with an ideal-gas equation of state,
$p=(\gamma-1)\rho e$, using $\gamma=1.4$.
The initial state is uniform with $v=0$ and $\rho=1$,
and internal energy initialized by a delta-like source at the origin normalized
to integrate to a total deposited energy of $0.25$.
The simulation is advanced to final time $T=0.8$.

\begin{figure}[pos=htbp]
\centerline{%
\begin{minipage}[c]{0.30\textwidth}%
\centering
\includegraphics[width=\linewidth]{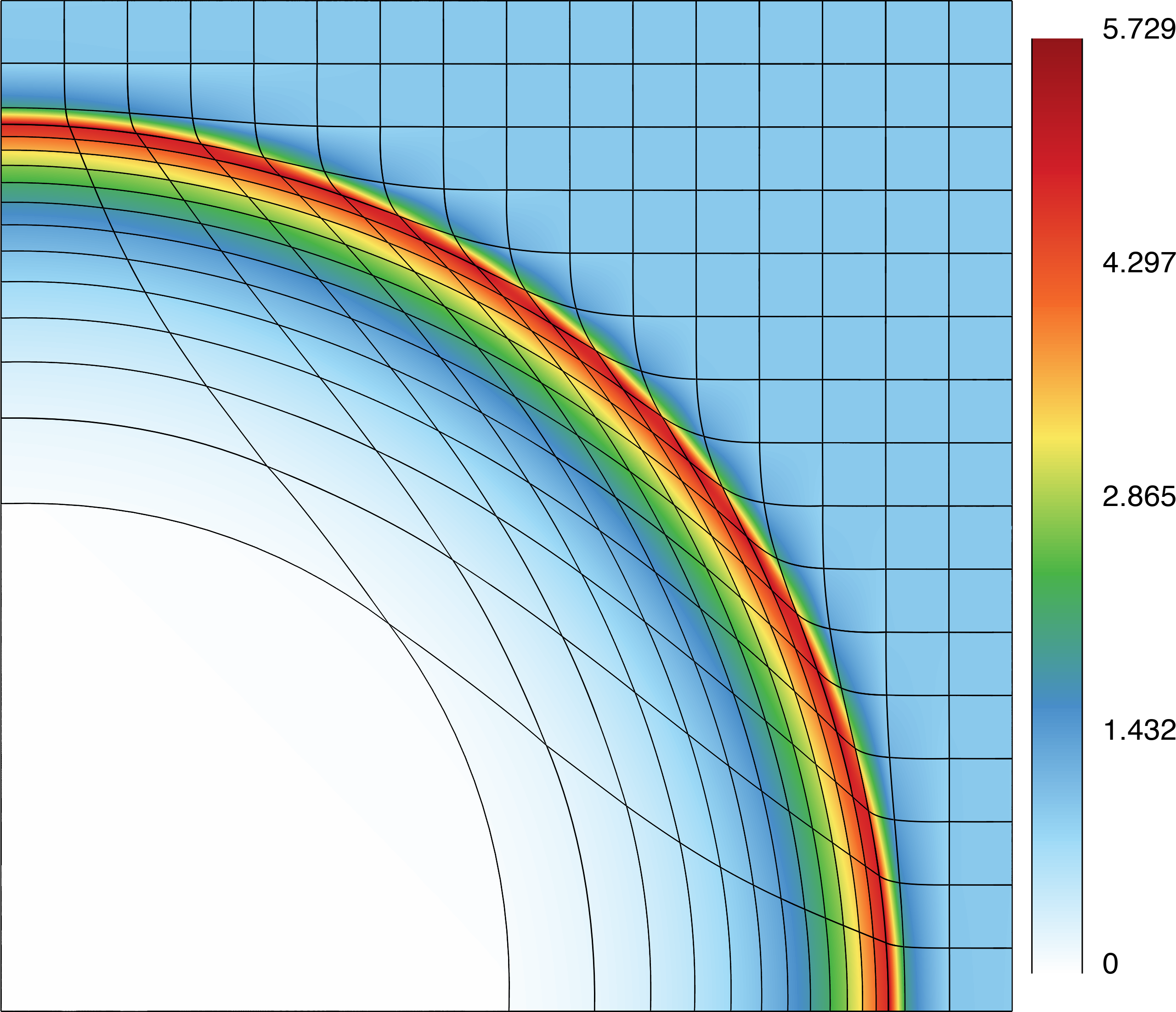}%
\end{minipage}%
\hspace{0.6em}%
\begin{minipage}[c]{0.30\textwidth}%
\centering
\includegraphics[width=\linewidth]{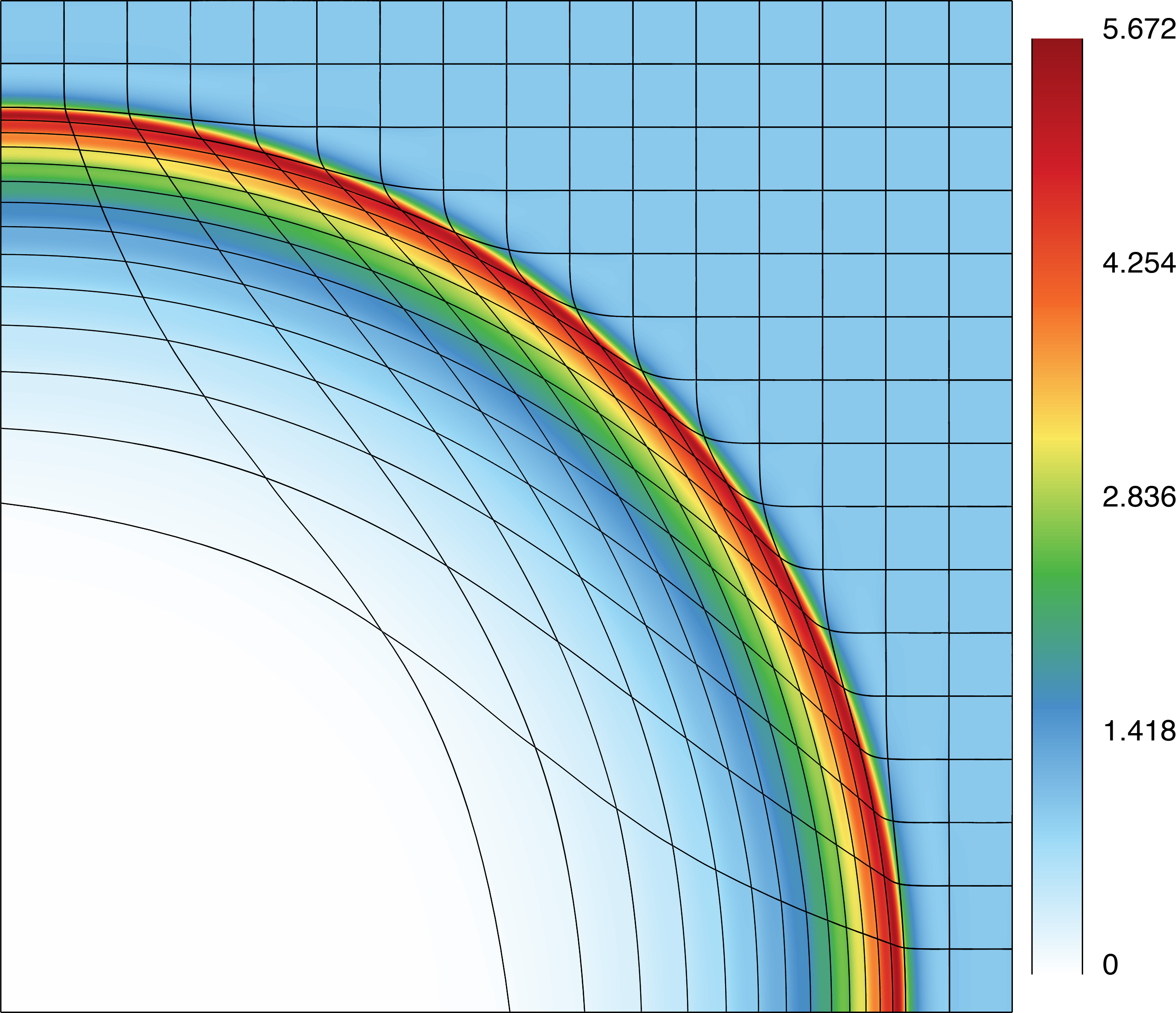}%
\end{minipage}%
\hspace{0.6em}%
\begin{minipage}[c]{0.40\textwidth}%
\centering
\includegraphics[width=\linewidth]{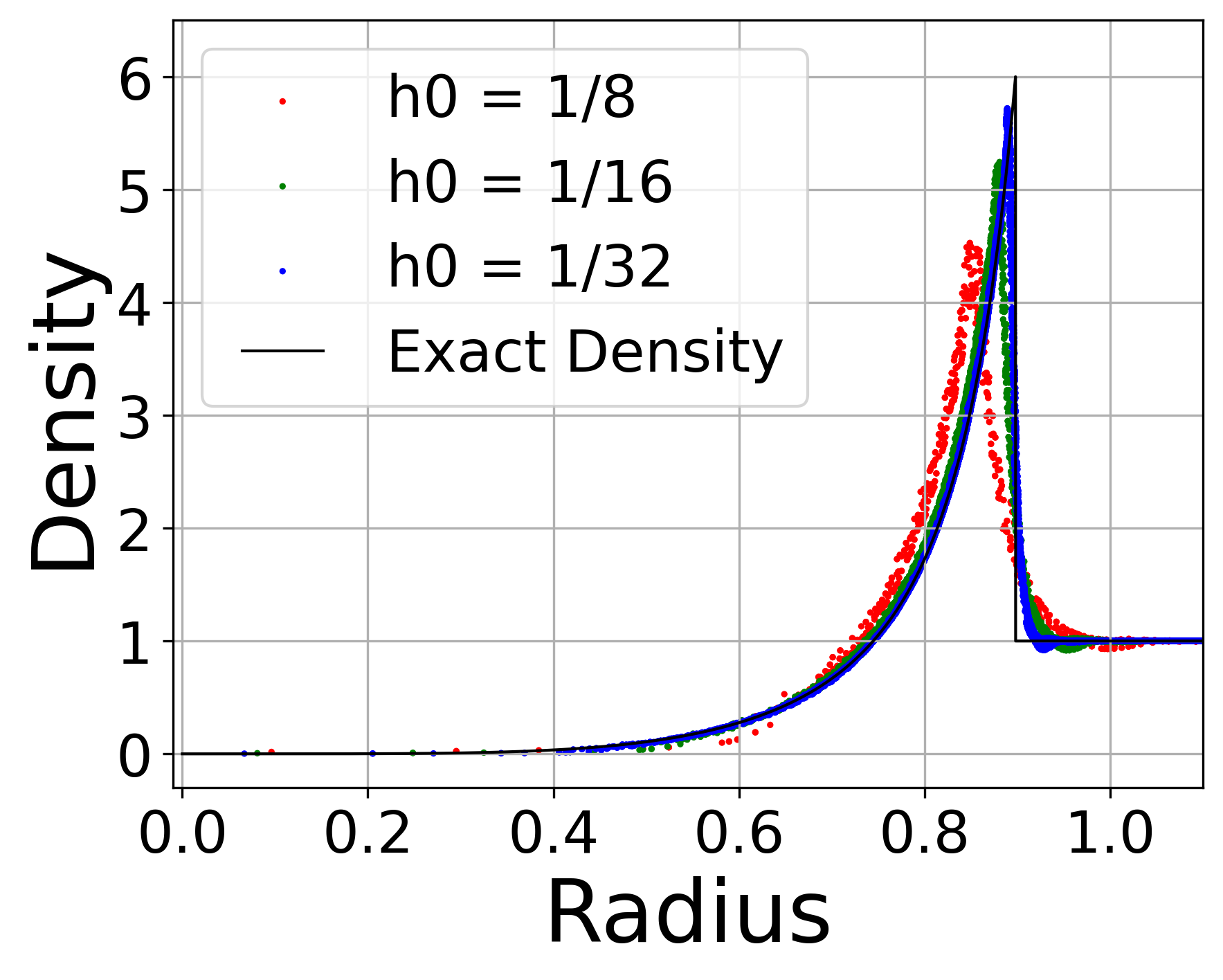}%
\end{minipage}%
}
\caption{2D Sedov blast density at final time $T=0.8$.
         Left: final result from the original explicit method.
         Middle: final result from the implicit simulation.
         Right: implicit scatter-plot results for three refinement
         levels compared to the exact solution.}
\label{fig_sedov_scatter}
\end{figure}

The left and middle panels of Figure~\ref{fig_sedov_scatter} show final density
fields computed on a $16 \times 16$ mesh with a Q3Q2 spatial discretization
(cubic kinematic space and quadratic thermodynamics): the original explicit
simulation uses RK2 with CFL 0.5 and requires 1157 time steps, while the
implicit run uses FS-CFL 2.0 and requires only 205 steps using the {\em implicit
midpoint} (IM) time stepping method.
The two results are very comparable, showing that the implicit method remains
accurate and robust under large time steps, with only slightly worse mesh
deformation.
The right panel shows scatter-plot results with implicit time stepping under
mesh refinement using a Q2Q1 + IM discretization. We see that the shock
position and overall profile converge to the expected exact Sedov solution.

To further demonstrate the applicability of the method in three dimensions,
in Figure~\ref{fig_3D} (left) we include a 3D Sedov blast wave result from a
Q2Q1 method with the same implicit formulation and viscosity treatment.
The simulation was run with a FS-CFL number of 2.0 and completed in 103 implicit
midpoint time steps while correctly capturing the radial shock propagation
and overall solution structure.

%---------------------

\subsection{Triple Point}
\label{sec_test_triple}

The triple point problem is a classic multi-material hydrodynamics benchmark
that produces rich interface dynamics and strong shock interactions.
This test is particularly well suited for demonstrating improved
accuracy-per-time-to-solution with implicit time integration, since the stable
time step of the explicit method drops drastically at later times, especially
for high-order simulations.
The setup of this problem is described in Section~8.5 of \cite{Dobrev2012}.

\begin{figure*}[pos=htbp]
\begin{center}
$\begin{array}{ccc}
\includegraphics[height=1.2in]{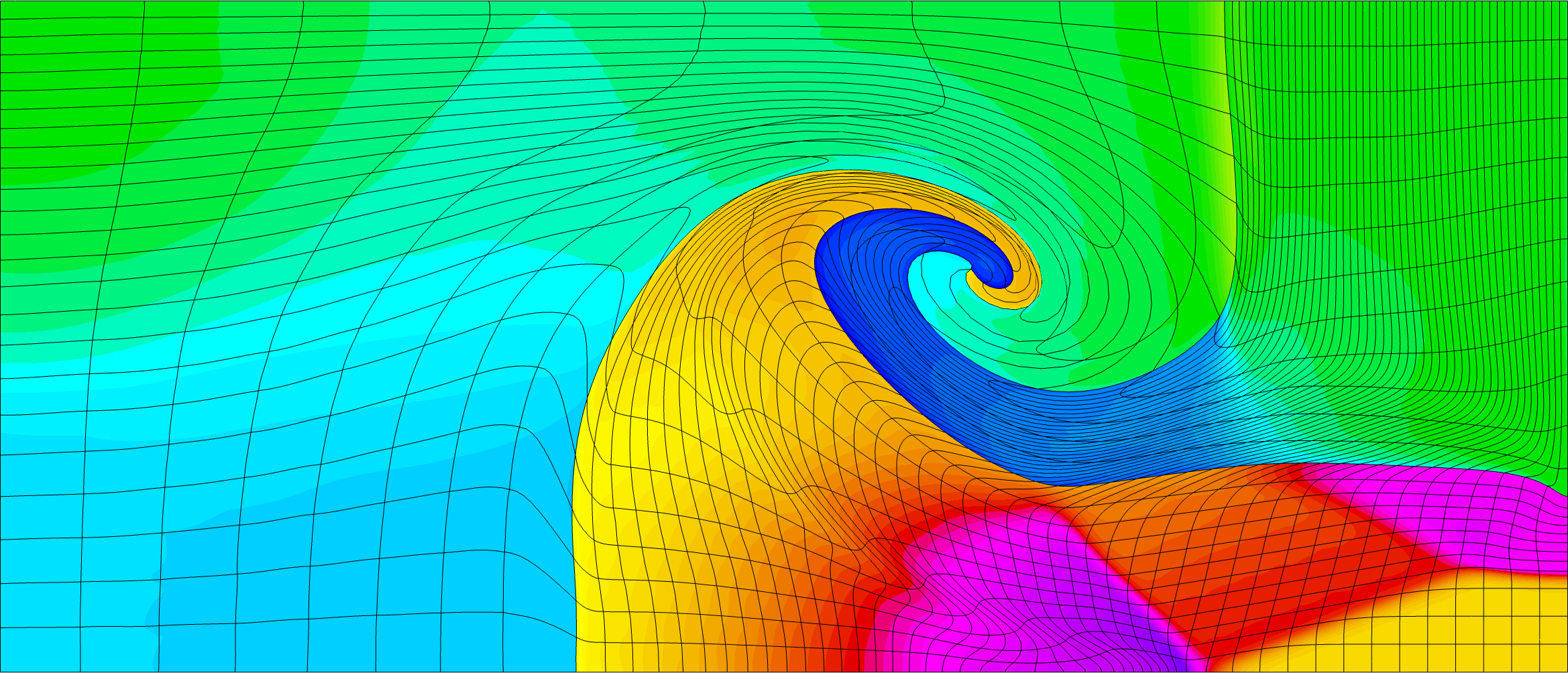} &
\includegraphics[height=1.2in]{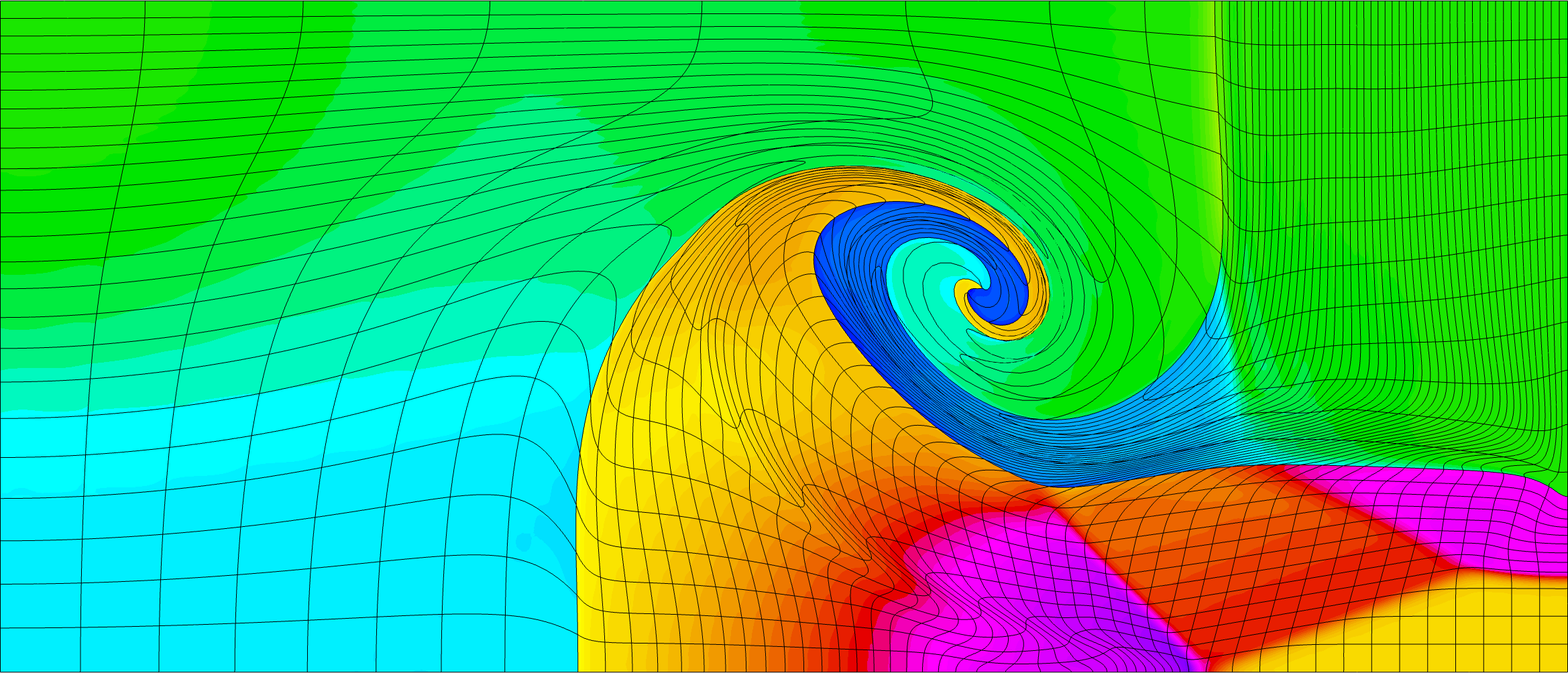} &
\includegraphics[height=1.2in]{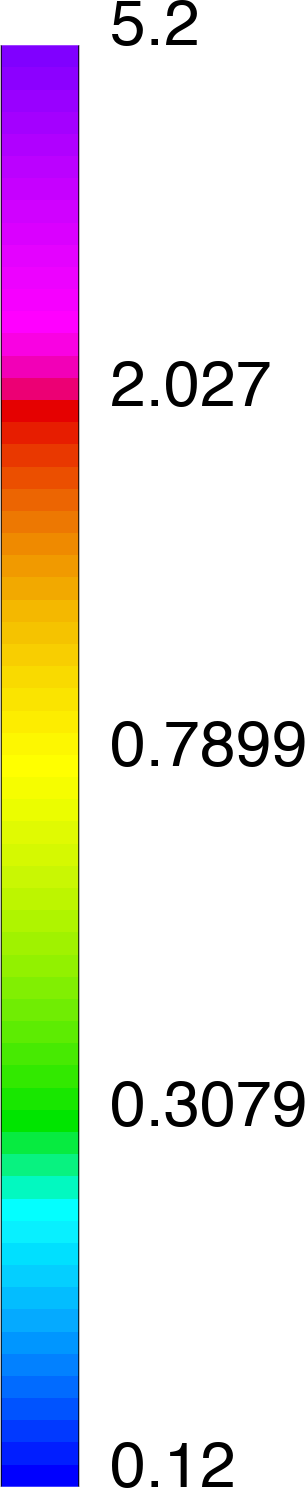} \\
\end{array}$
\end{center}
\caption{2D Triple Point density field using a Q4Q3 spatial discretization.
         Left: explicit run (232{,}649 time steps, 4{,}216 seconds).
         Right: implicit run (191 time steps, 292 seconds).}
\label{fig:2D-triple-point}
\end{figure*}

We consider a high-order Q4Q3 + RK2 discretization.
The explicit run required $236{,}607$ time steps and $10{,}710$
seconds to reach the final time.
In contrast, using the {\em implicit midpoint} (IM) time stepping method with
the same FS-CFL value of
$0.5$ for the initial time step, as described above, the implicit formulation
reached the final time in $191$ time steps and $349$ seconds.
Both runs were executed on the same computational resource:
4 MPI tasks on a personal laptop.
The corresponding density fields are shown in Figure \ref{fig:2D-triple-point}.
The resulting density fields indicate improved resolution of the roll-up
structures in the implicit case, despite the substantially larger time steps.
These results highlight the substantial potential of the implicit approach to
deliver better accuracy per unit runtime on problems where explicit stability
constraints become increasingly severe; additional discussion is provided in
Section~\ref{sec_accuracy_time}.

\begin{figure*}[pos=htbp]
\begin{center}
$\begin{array}{ccc}
\includegraphics[width=0.4\textwidth]{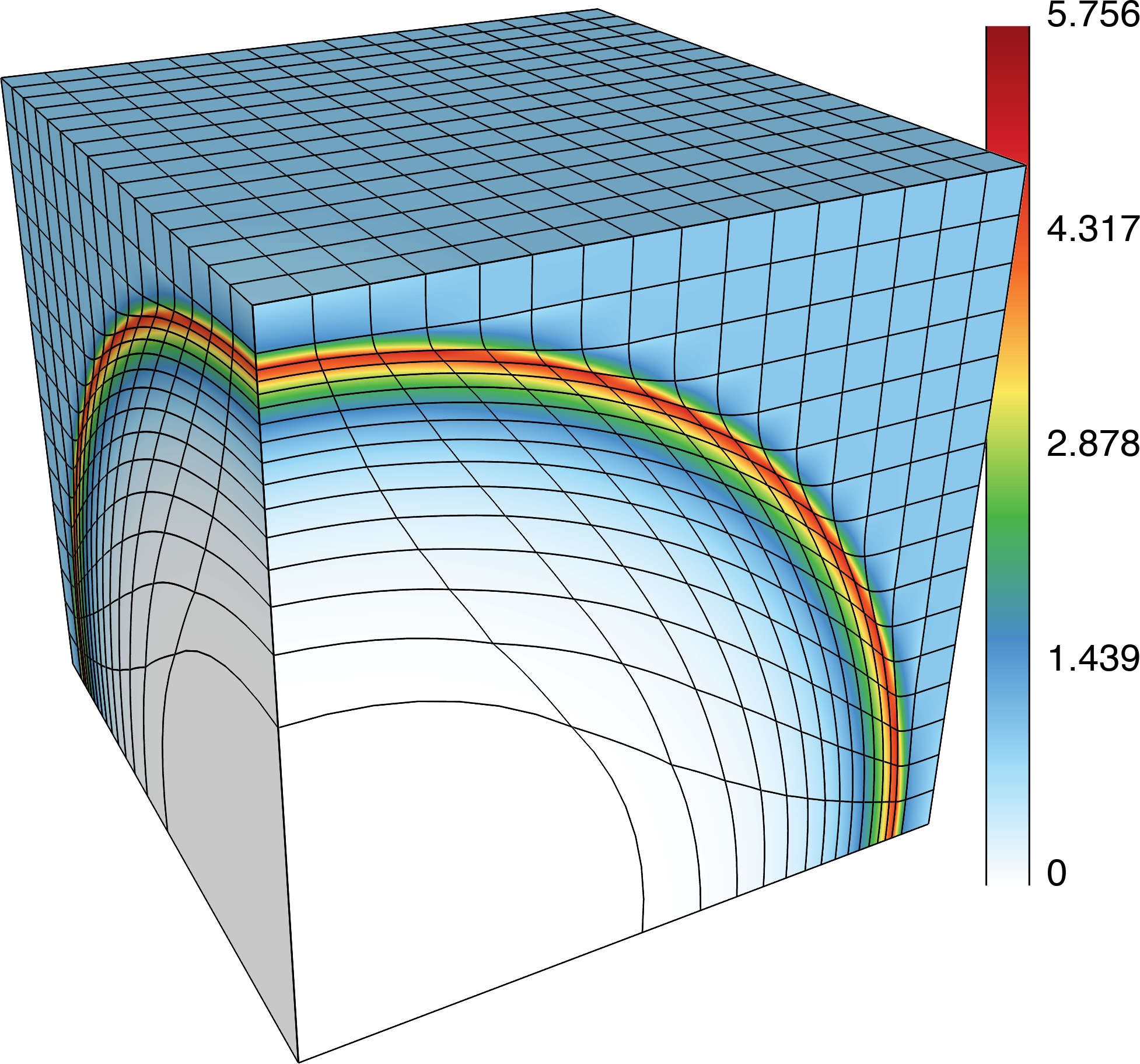} ~~
\includegraphics[width=0.5\textwidth]{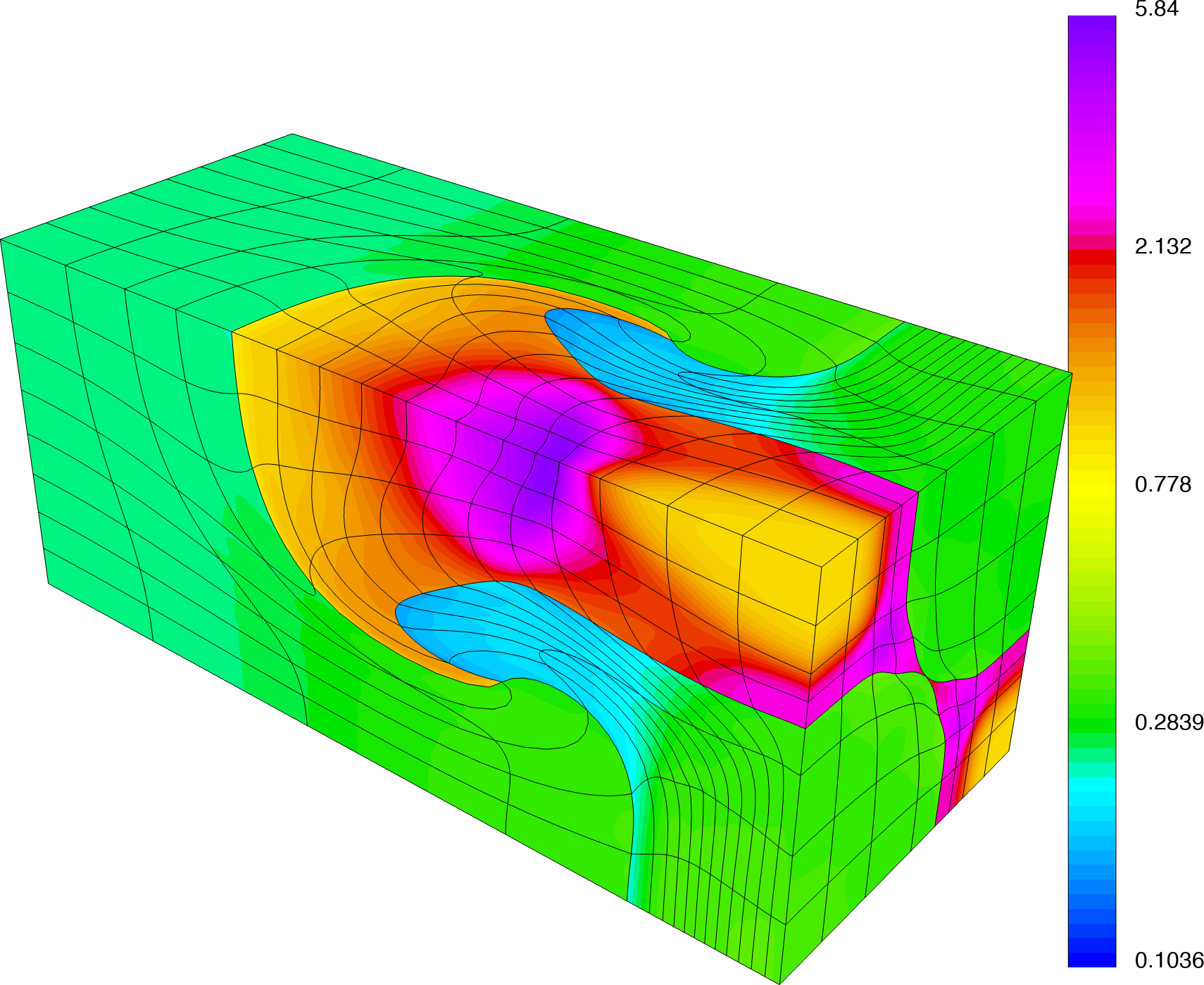}
\end{array}$
\end{center}
\caption{3D implicit simulation results for the Sedov Blast (left) and
         Triple Point (right) problems.}
\label{fig_3D}
\end{figure*}

Similarly to the previous test cases, in Figure~\ref{fig_3D} (right) we show a
Q3Q2 3D Triple Point result that demonstrates that the implicit formulation
extends robustly to three dimensions without modification of the overall approach.
The simulation was run to time $t=5$ with a FS-CFL number of $0.5$ and completed
in 74 implicit midpoint time steps while correctly capturing the expected
large-scale flow structures and interface evolution.

%---------------------

\subsection{Accuracy-per-Time Discussion}
\label{sec_accuracy_time}

One of the motivations for implicit hydrodynamics methods is the possibility of
improving the accuracy obtained per unit of computational time by taking
significantly larger time steps than explicit methods.
The results presented in this work demonstrate that such improvements are
indeed achievable in practice for certain problems and parameter regimes.
In particular, the implicit discretization was able to outperform the explicit
method in overall time-to-solution while achieving better accuracy on
problems where the explicit time step becomes strongly restricted, e.g.\ due to
significant mesh stretching.
At the same time, these gains are not systematic and
depend on several interacting factors.

First, the efficiency of the implicit method depends strongly on the quality of
the nonlinear solver and preconditioner.
The current implementation uses Newton iterations with Krylov linear solves and
a block-diagonal AMG preconditioner applied to partially assembled operators.
While this approach already provides competitive performance, the nonlinear and
linear solver components are far from fully optimized.
Improved preconditioning strategies, more advanced nonlinear globalization
techniques, matrix-free smoothers, better reuse of Jacobian information, and
architecture-specific optimizations could significantly alter the balance
between explicit and implicit runtimes.

Second, the comparison depends on the artificial viscosity
formulation used in the two methods.
The implicit discretization employs a smoother viscosity model suitable for
Newton-based nonlinear solves, whereas the explicit formulation relies on
sharper shock-detection mechanisms.
As a result, there is no exact one-to-one correspondence between the
dissipative behavior of the explicit and implicit schemes,
especially when the implicit time step becomes large compared to the explicit
one.
The effective sharpness of shocks and small-scale features therefore depend not
only on the time step size, but also on the viscosity regularization required
for robust nonlinear convergence.

Third, the relative performance depends strongly on the dynamics of the
underlying problem.
For example, in the Triple Point problem, the explicit method
eventually becomes constrained by very small stable time steps,
while the implicit method remains stable with significantly larger
time steps and achieves better accuracy per unit runtime.
In contrast, the Sedov blast wave problem exhibits extremely strong localized
mesh deformation and rapidly evolving shock dynamics that effectively impose
geometric restrictions on the time step even in the implicit setting.
In such regimes, the ability of implicit methods to take arbitrarily large time
steps becomes severely limited, and explicit methods may remain preferable.

Several additional factors also influence the practical efficiency tradeoff
between explicit and implicit methods:
\begin{itemize}
\item the differences in time step control strategies between the explicit and
      implicit formulations, including the frequency of repeated or rejected
      time steps and the magnitude of time step reductions,
\item the scalability, memory movement, and parallel efficiency of the
      underlying linear algebra kernels,
\item the robustness of the nonlinear solver in the presence of strong shocks,
      mesh distortion, and rapidly evolving flow features,
\item the order of the spatial and temporal discretizations,
\item and the desired accuracy tolerances of the nonlinear iterations.
\end{itemize}

%-------------------------------------------------

\section{Conclusion}
\label{sec_concl}

We have developed an implicit Lagrangian hydrodynamics capability that builds
directly on the existing high-order finite element discretization described
in \cite{Dobrev2012} and demonstrated using the Laghos miniapp (built on MFEM), 
without changing the underlying physics formulation.
By rewriting the original pointwise kernels in MFEM's \dfem interface and
leveraging Enzyme-based automatic differentiation, we obtain the required
derivatives, including Jacobian actions, automatically, enabling a
Newton-Krylov solve with a matrix-free or assembled Jacobian operator.

The numerical experiments demonstrate the expected high-order convergence on a
smooth Taylor-Green vortex problem, correct behavior on a strong-shock Sedov
blast, and accuracy-per-time improvements on the Triple Point
benchmark where the explicit method becomes severely limited by late-time
stability constraints for high-order discretizations.
The same formulation extends directly to three dimensions and
produces robust results without modification of the overall approach.
Overall, the results indicate that implicit hydrodynamics can provide
substantial improvements in accuracy-per-time-to-solution for appropriate
classes of problems, but these gains are inherently problem-dependent and
closely tied to solver robustness, dissipation modeling, and implementation quality.

Future research directions include: extending the implicit capability to additional physics
models and improving the nonlinear and linear solvers and preconditioners,
including more effective matrix-free and physics-based preconditioning strategies,
and adapting the hyper-viscosity limiting approach for differentiability.
Finally, the availability of accurate gradients through \dfem opens the door to
new simulation workflows beyond implicit time integration, including adjoint
optimization, sensitivity analysis, and parameter calibration.

\paragraph{License Notice}
This manuscript has been authored by Lawrence Livermore National Security,
LLC under Contract No. DE-AC52-07NA27344 with the U.S. Department of Energy.
The United States Government retains, and the publisher, by accepting the 
article for publication, acknowledges that the United States Government retains
a non-exclusive, paid-up, irrevocable, world-wide license to publish or
reproduce the published form of this manuscript, or allow others to do so,
for United States Government purposes.

\bibliographystyle{cas-model2-names}
\bibliography{bib}

%-------------------------------------------------

\end{document}